\documentclass[reprint,aps,prd,twocolumn,preprintnumbers,
superscriptaddress,showpacs,nofootinbib]{revtex4-1}
\usepackage{plain}
\usepackage{graphicx}
\usepackage{amsmath,amsfonts,amssymb}
\usepackage{epsfig}
\usepackage[hyperindex,breaklinks]{hyperref} 
\usepackage{latexsym}
\usepackage{graphicx}
\usepackage{amsmath}
\usepackage{amsfonts}   
\usepackage{amssymb}    
\usepackage{float}
\usepackage{bm}
\usepackage{slashed}
\usepackage{hyperref}
\usepackage{color}
\usepackage{breakurl}

\usepackage{color}
\definecolor{myred}{rgb}{0.6,0,0} 
\definecolor{myblue}{rgb}{0,0.2,0.4}
\definecolor{mygreen}{rgb}{0,0.9,0.1}
\definecolor{hc}{rgb}{.9,0.1,0.7}
\definecolor{hcout}{rgb}{.9,0.7,0.9}
\definecolor{Orange}{rgb}{1.,0.65,0.}
\def\lsim{\raise0.3ex\hbox{$\;<$\kern-0.75em\raise-1.1ex\hbox{$\sim\;$}}}
\def\gsim{\raise0.3ex\hbox{$\;>$\kern-0.75em\raise-1.1ex\hbox{$\sim\;$}}}
\def\mdd{m_{\Delta^{\pm\pm}}}
\def\dc{\Delta^{\pm\pm}}

\newcommand{\met}{\ensuremath{\not\!\!E_T}\xspace}
\newcommand{\be}{\begin{equation}}
\newcommand{\ee}{\end{equation}}
\newcommand{\bea}{\begin{eqnarray}}
\newcommand{\eea}{\end{eqnarray}}

\newcommand{\newc}{\newcommand}
\newc{\bi}{\begin{itemize}}
\newc{\ei}{\end{itemize}}

\newc{\ra}{\rightarrow}
\newc{\sq}   {\mbox{$\wt{q}$}}
\newc{\msq}  {\mbox{$m_{\sq}$}}
\newc{\gl}   {\mbox{$\wt{g}$}}
\newc{\mgl}  {\mbox{$m_{\gl}$}}
\def \met  {\mbox{${E\!\!\!\!/_T}$}}
\newc{\wt}{\widetilde}

\newc{\ifb}{\mbox{${\rm fb}^{-1}$}}

\newc{\del}{\delta}

\begin{document}

\preprint{CPHT-RR054.1115, LPT-Orsay-15-89, HRI-RECAPP-2015-016}
\title{Reconciling {\boldmath$(g-2)_\mu$} and charged lepton flavour violating 
processes through a doubly charged scalar}

\author{Joydeep Chakrabortty} \affiliation{Department of Physics,   
Indian Institute of Technology, Kanpur-208016, India}

\author{Pradipta Ghosh} \affiliation{Laboratoire de Physique Th\'eorique, CNRS\footnote{UMR  8627}, 
Univ. Paris-Sud, Universit\'e Paris-Saclay, 
91405 Orsay, France 
Centre de Physique Th\'eorique, Ecole polytechnique, CNRS\footnote{UMR 7644}, 
Universit\'e Paris-Saclay,
91128 Palaiseau, France}

\author{Subhadeep Mondal } \affiliation{Regional Centre for Accelerator-based Particle Physics,
Harish-Chandra Research Institute, Jhunsi, Allahabad - 211019, India}

\author{Tripurari Srivastava} \affiliation{Department of Physics,   
Indian Institute of Technology, Kanpur-208016, India}



\vskip 0.3cm 
\begin{abstract}
The scalar particle discovered at the Large Hadron Collider (LHC)
has properties very similar to that of a standard model (SM) Higgs boson.
Limited experimental knowledge of its model origin, as of now,
however, does not rule out the possibility of accommodating this new
particle into a  beyond  the SM (BSM) framework. A few of these schemes
suggest that the observed scalar is just the {\it lightest} candidate
of an enriched  sector with several other heavier states  {\it awaiting
to be detected}. Such models with {\it nonminimal} scalar sector
also accommodate other neutral and electrically charged (singly, doubly, triply, etc.)
component fields as prescribed by the specific model. Depending on the mass
and electric charge, these new states can 
produce potential signatures at  colliders as well as in low-energy
experiments. 
The presence of a doubly charged scalar, when accompanied by other neutral
or charged scalar(s), can also generate neutrino masses. Adopting the 
second scenario, e.g., Babu-Zee construction, constraints from neutrino
physics have been effaced in this study.
Here, we investigate a few phenomenological 
consequences of a uncoloured  doubly charged scalar which  
couples to the charged leptons as well as gauge bosons. Restricting ourselves
in the regime of conserved charged-parity (CP), we assume 
only a few nonzero Yukawa couplings ($y_{\mu \ell}$, where $\ell=e,\mu,\tau$) 
between the doubly charged scalar and the charged leptons. Our choices
allow the doubly charged scalar to impinge low-energy processes like
anomalous magnetic moment of muon and a few possible charged lepton flavour 
violating (CLFV) processes. These same Yukawa couplings are also instrumental in
producing same-sign dilepton signatures at the LHC. 
 In this article we examine 
the impact of individual  contributions from the diagonal and off-diagonal Yukawa 
couplings in the light of muon $(g-2)$ excess. Subsequently, we use 
the derived information to inquire the possible
CLFV processes and finally the collider signals from the decay
of a doubly charged scalar. Our simplified analyses, depending
on the mass of  doubly charged scalar, provide a
good estimate for the magnitude of the concerned Yukawa couplings.
Our findings would appear resourceful to test the phenomenological
significance of a doubly charged scalar by using complementary information
from muon $(g-2)$, CLFV and the collider experiments.
\end{abstract}

\maketitle
\section{Introduction}\label{Introduction}
Discovery of a new scalar \cite{Aad:2012tfa,Chatrchyan:2012ufa} has 
already proclaimed the success of  LHC. This scalar  has properties \cite{Aad:2015zhl,atlcms} 
quite identical to 
that of the SM Higgs boson, the only fundamental scalar within the SM framework. 
In the SM, the Higgs field emerges from an $SU(2)$ complex scalar doublet. 
A complete knowledge of
the SM Higgs sector would require (i) measurements of the vacuum expectation value (VEV)
 acquired by the electrically neutral CP-even component of the aforementioned
complex scalar doublet, (ii) the Higgs boson mass and (iii) the Higgs self coupling.
At present, we have already probed the VEV of the SM through experimental 
measurements \cite{Agashe:2014kda} and, have estimated the mass of a
Higgs-like scalar boson at the LHC \cite{Aad:2015zhl}. Thus, it remains to examine the only
remaining parameter of the SM-Higgs sector, namely, the self-coupling.
Unfortunately, the experimental sensitivity for the latter is very poor
at the LHC and one perhaps needs to wait for the future colliders \cite{Baer:2013cma}.
Hence, the possibility of having a Higgs-like scalar from BSM theories
is certainly not redundant till  date, especially when several other observations already 
ask for such an extension, e.g., nonzero neutrino masses and mixing \cite{Tortola:2012te,
Fogli:2012ua,GonzalezGarcia:2012sz}. Furthermore, mass of the newly 
discovered scalar \cite{Aad:2015zhl} and the top-quark mass \cite{Agashe:2014kda}
strongly prefer the presence of one or more {\it BSM} scalars in the theory 
before $10^{9}-10^{11}$ GeV \cite{Sher:1988mj,EliasMiro:2012ay,Alekhin:2012py,Buttazzo:2013uya}\footnote{Some
counter arguments also exist in this connection, as addressed in Ref. \cite{Jegerlehner:2013cta}.} 
(see also Refs.~\cite{Djouadi:2005gi,Djouadi:2005gj} for review). 
Introduction of these new scalars assures stability of the SM-Higgs potential
up to the Planck scale. Combining these observations,
an extension of the SM Higgs sector seems rather plausible. For example, one can add extra
scalar states which are encapsulated in different multiplets guided by the gauge symmetry 
and/or pattern of the symmetry breaking.
A plethora of analyses \cite{Pati:1974yy,
Mohapatra:1974hk,Mohapatra:1974gc,Senjanovic:1975rk,Konetschny:1977bn,Senjanovic:1978ev,
Magg:1980ut,Schechter:1980gr,Cheng:1980qt,Zee:1980ai,Lazarides:1980nt,Mohapatra:1980yp,Zee:1985rj,
Georgi:1985nv,Chanowitz:1985ug,Zee:1985id,Babu:1988ki,McDonald:1993ex,
Bento:2000ah,Burgess:2000yq,Babu:2002uu,Davoudiasl:2004be,
Schabinger:2005ei,Cirelli:2005uq,Kusenko:2006rh,Chen:2006vn,O'Connell:2006wi,
BahatTreidel:2006kx,Chen:2007dc,Gogoladze:2008gf,Barger:2008jx,Hambye:2009pw,Dawson:2009yx,Babu:2009aq,
Gonderinger:2009jp,Aoki:2011yk,delAguila:2011gr,Lebedev:2012zw,
EliasMiro:2012ay,Chun:2012jw,Chao:2012mx,Dev:2013ff,Barry:2013xxa,Cline:2013gha,
Chakrabortty:2013zja,Chakrabortty:2013mha,King:2014uha,Okada:2014qsa,Costa:2014qga,Martin-Lozano:2015dja,
Falkowski:2015iwa,Bonilla:2015kna,Okada:2015hia,Das:2015nwk,Bambhaniya:2015ipg} already exists 
in this connection where additional scalar mulitiplets  are introduced to solve different shortcomings 
of the SM, like stability of the scalar potential
up to the Planck scale, dark-matter, neutrino masses and mixing, etc.

These  BSM scalar multiplets in general contain not only the electrically neutral fields but the 
charged (singly, doubly, triply, etc.) ones also. Phenomenology of these states may 
be constrained from the electroweak precision tests \cite{Agashe:2014kda}, e.g., see 
Refs. \cite{Blank:1997qa,Melfo:2011nx,Chun:2012jw,Bonilla:2015eha} for an 
extension with $SU(2)$ triplet Higgs. The presence
of charged scalars,  depending on the structure of the associated multiplet, at the same time
can produce novel signals at the collider experiments, e.g., same-sign multileptons.
Several analyses \cite{Gunion:1996pq,Chakrabarti:1998qy,Chun:2003ej,Muhlleitner:2003me,Akeroyd:2005gt,Chen:2006vn,
Han:2007bk,Garayoa:2007fw,Kadastik:2007yd,Akeroyd:2007zv,Perez:2008ha,delAguila:2008cj,
Akeroyd:2009hb,Maiezza:2010ic,Akeroyd:2010ip,Tello:2010am,Rentala:2011mr,Aoki:2011yk,
Akeroyd:2011zza,Melfo:2011nx,Aoki:2011pz,Akeroyd:2012nd,
Chakrabortty:2012pp,Das:2012ii,Chun:2012zu,Chen:2012vm,
Kanemura:2013vxa,Bambhaniya:2013yca,delAguila:2013yaa,Chun:2013vma,delAguila:2013mia,
Bambhaniya:2013wza,Dutta:2014dba,Kanemura:2014bqa,Kanemura:2014goa,
Bambhaniya:2014cia,Deppisch:2014zta,Han:2015hba,Bambhaniya:2015wna} are already performed in this direction, 
including experimental ones \cite{Acton:1992zp,Abbiendi:2001cr,Abdallah:2002qj,Abbiendi:2003pr,
Achard:2003mv,Abazov:2004au,Acosta:2004uj,Azuelos:2004mwa,
Rommerskirchen:2007jv,Hektor:2007uu,Abazov:2008ab,Aaltonen:2008ip,Abazov:2011xx,Aaltonen:2011rta,
Chatrchyan:2012ya,ATLAS:2012hi,ATLAS:2014kca}.
These charged scalars, apart from atypical LHC signatures, can also contribute to
a class of low-energy phenomena that lead to lepton number as well as flavour violating
processes. Such processes include CLFV (e.g. $\mu \to e \gamma,\,$ 
$\mu$ to $e$ conversion in atomic nuclei etc.) \cite{Petcov:1982en,Leontaris:1985qc,Bernabeu:1985na,
Bilenky:1987ty,Swartz:1989qz,Chun:2003ej,Kakizaki:2003jk,Cirigliano:2004mv,
Cirigliano:2004tc,Abada:2007ux,Fukuyama:2009xk,Ren:2011mh,Chakrabortty:2012vp,
Dinh:2012bp,Das:2012ii,Barry:2013xxa,Nayak:2013dza,King:2014uha,Borah:2014bda,Deppisch:2014zta}, 
neutrinoless double beta decay ($0\nu\beta\beta$) \cite{Mohapatra:1981pm,
Haxton:1982ff,Wolfenstein:1982bf,Hirsch:1996qw,Cirigliano:2004tc,Chen:2006vn,
Tello:2010am,delAguila:2011gr,Chakrabortty:2012mh,Barry:2013xxa,Dev:2014xea,King:2014uha,Deppisch:2014zta}, 
rare meson decays (e.g., $M \to M' \ell_i \ell_j, \, 
M \to M' \ell_i \ell_j \ell_m \ell_n$) \cite{Picciotto:1997tk,Ma:2009fi,Quintero:2012jy,
Bambhaniya:2015nea}, muon $(g-2)$ \cite{Fukuyama:2009xk,Freitas:2014pua} etc. 
Some of these processes, e.g., $0\nu\beta\beta$, rare-meson
decays, etc. have one thing in common, i.e.,  they violate lepton number by 
$2$ units which is the characteristic of a doubly-charged scalar\footnote{Presence
of a Majorana fermion, for example right-handed neutrino, can also serve the 
same purpose, see Ref.~\cite{Atre:2009rg} and references therein.}. 
The same doubly charged scalar can also participate
in the CLFV processes and muon $(g-2)$. Several investigations, as aforesaid, 
do already exist concerning various phenomenological
aspects of a doubly charged scalar. A dedicated entangled phenomenological
inspection of the doubly-charged scalars, in the context of collider 
and low-energy experiments at the same time, however, still remains somewhat 
incomplete. This is exactly what we plan to do here and the current
article is the first step toward a complete investigation.
In passing we note  that the other part of   multiplets, i.e., the neutral scalar 
states can also show their own distinctive signals. For example, 
if these BSM neutral scalars  are light, they can affect  
the SM-Higgs decay phenomenology  through mixing. 
Phenomenological implications of additional neutral scalars are however, beyond the 
theme of this article and will not be addressed further.

We initiate our investigation with the discrepancy in anomalous magnetic moment 
of muon $\Delta a_\mu=a^{exp}_\mu-a^{th}_\mu$, which can be 
explained well in the presence of a doubly-charged scalar, $\Delta^{\pm\pm}$. Subsequently, we use 
this information to constrain {\it only the most relevant} associated
Yukawa couplings that connect the doubly charged scalar with the charged leptons,
i.e., $y_{\mu\ell}$ with $\ell=e,\,\mu,\,\tau$.
In the next step, we investigate the allowed relevant CLFV processes
 in the presence of {\it the same set} of Yukawa couplings.
At this level we scrutinize a new set of constraints on that 
{\it same set} of Yukawa couplings from the experimental limits on
different CLFV processes. Finally, we explore the collider
signals of a doubly charged scalar that appear feasible with
the {\it chosen set} of Yukawa couplings and, are in agreement with
the experimental constraints of $(g-2)_\mu$ and the relevant CLFV processes.
However, like the existing literature we do not work in the context of
any specific model. Rather, we parametrize the {\it unknown} model of
doubly charged scalar in terms of a {\it few relevant} Yukawa couplings and the mass
of the doubly charged scalar $(\mdd)$ that are
resourceful to probe the existence of a doubly charged scalar experimentally.
As a first attempt, we also stick to the regime of conserved CP. It is also
important to emphasise that we focus on the range of\footnote{The exclusion limit depends 
on the leptonic decay branching fractions of $\dc$ and one can 
safely take $m_{\Delta^{\pm \pm}} \geq 400$ GeV.} 
$400~{\rm GeV}$ \cite{ATLAS:2014kca} $\lsim \mdd \lsim 1000~{\rm GeV}$ which is well 
accessible during run-II of the LHC\footnote{One can always
consider a large value for $\mdd$ to suppress the CLFV processes. However, a
heavier $\dc$ would result in a smaller production cross-section at the LHC and thereby 
ends in smaller number of signal events.}.
Thus, in a nutshell, in this work we explore the possible correlations among 
$\mdd$ and $y_{\mu\ell}$ as well as between different
$y_{\mu\ell}$ in the context of (i) $(g-2)_\mu$ and (ii)
a few CLFV processes. Thereafter, we use 
these information to study the possible $\Delta^{\pm\pm}\to \ell^\pm_{\alpha} \ell^\pm_{\beta}$
processes (i.e., same-sign dileptons) at the LHC.

It remains to mention one more important aspect associated with a $\dc$, i.e., the generation 
of nonzero neutrino masses and mixing. Accommodating massive neutrinos in the presence 
of a $\dc$ depends on the chosen theory framework which we will discuss later 
in Sec. \ref{Scenario}. Models of these kinds typically contain additional scalars
(neutral, charged or both, depending on the concerned model) and a larger set 
of Yukawa couplings. This nonminimal set of Yukawa couplings (compared to $y_{\mu\ell}$)
is essential to accommodate massive neutrinos simultaneously with
an explanation for the $(g-2)_\mu$ anomaly, in the presence of a few CLFV processes.
In the context of a minimal model (described later), we 
observed that the set of constraints on the relevant Yukawa couplings coming 
from the anomalous $(g-2)_\mu$ and nonobservation of the CLFV processes
is rather independent to the ones required to satisfy the observed
three flavour global neutrino data \cite{Tortola:2012te,Fogli:2012ua,GonzalezGarcia:2012sz}.

The paper is organised as follows, after the introduction we present
a concise description of the underlying theoretical framework in Sec. \ref{Scenario}.
Analytical expressions for the muon $(g-2)$ and a few possible
CLFV processes are given in Sec. \ref{gm2} and Sec. \ref{clfv}, respectively.
We present  results of our numerical analyses on $\Delta a_\mu$ and 
the allowed CLFV processes in Sec. \ref{result}. Collider phenomenology
of the $\dc\to\ell^\pm_{\alpha}\ell_{\beta}^\pm$ processes, following findings of
the previous section is addressed in Sec. \ref{collider}. Our conclusions
are given in Sec. \ref{conclusion}. Finally, a detail 
computation of the anomalous magnetic moment of muon 
through a doubly charged scalar is relegated in the appendix.

\section{The theory framework}\label{Scenario}

The presence of a doubly charged scalar is possible in various
representations, for example (i) an $SU(2)$ triplet\footnote{It has 
been mentioned in Refs.~\cite{Fukuyama:2009xk,Kohda:2012sr}, 
that in the context of neutrino mass generation
using a scalar $SU(2)$ triplet, i.e., Type-II seesaw,
it is normally difficult to generate additive contributions
to muon $(g-2)$ from $(y^{\dagger} y)_{\mu \mu}$. One can nevertheless,
generate an additive contribution to $(g-2)_\mu$
with $(y^2)_{\mu \mu}$ or $(y_{\mu \mu})^2$ \cite{Fukuyama:2009xk}.} 
$\delta^T \equiv (\delta^{++},\,\delta^+, \delta^0)$
with hypercharge, $Y=1$ \cite{Konetschny:1977bn,Senjanovic:1978ev,
Magg:1980ut,Schechter:1980gr,Cheng:1980qt}, (ii) an $SU(2)$ singlet $\kappa^{++}$
with $Y=2$ \cite{Babu:1988ki,delAguila:2011gr,Gustafsson:2012vj},
(iii) left-right symmetric model \cite{Pati:1974yy,
Mohapatra:1974hk,Mohapatra:1974gc,Senjanovic:1975rk},
(iv) a quadruplet $\Sigma^T \equiv (\Sigma^{++},\,\Sigma^{+},\,\Sigma^{0},\,\Sigma^{-})$
with $Y=1/2$ \cite{Ren:2011mh}, (v) another doublet $\chi^T \equiv (\chi^{++},\,\chi^+)$ with $Y=3/2$ 
\cite{Gunion:1996pq,Rentala:2011mr,Aoki:2011yk,Yagyu:2013kva},
(vi)  a quintuplet $\Omega^T \equiv (\Omega^{++},\,\Omega^{+},\,\Omega^{0},\,\Omega^{-},\,\Omega^{--})$
with $Y=0$ etc. The mulitiplets $\Sigma$ and $\chi$ give rise to dimension-five 
while $\Omega$ produces dimension-six neutrino mass operators through their respective interactions with
the SM fields. 
It is worth mentioning that a doubly charged scalar can also appear in a quintuplet 
with $Y>0$ or in multiplets with larger isospins \cite{Babu:2009aq,
Picek:2009is,Kumericki:2011hf,Ren:2011mh,Bambhaniya:2013yca,Earl:2013fpa}.

Phenomenological aspects of these multiplets are constrained from the electroweak precision 
observables \cite{Georgi:1985nv,Chanowitz:1985ug,Blank:1997qa,Melfo:2011nx,Hally:2012pu,Chun:2012jw,
Hisano:2013sn,Kanemura:2013mc,Earl:2013jsa,Bonilla:2015eha}, especially
if the multiplet contains an electrically neutral component which develops a VEV. 
For the simplicity of analysis we however, focus solely
on the doubly charged scalar without caring about the rest of multiplet
members. This approach helps us to pin down the precise contributions 
from the doubly charged scalar.
Furthermore, we assume negligible to vanishingly small
VEV  for the neutral scalar member of the associated multiplet, if any. The latter 
choice not only protects the $\rho$-parameter \cite{Agashe:2014kda}, but also
guarantees the absence and(or) severe suppression of some of the couplings,
e.g., $\dc W^\mp W^\mp$.

We are now ready to write down the relevant terms for
our analysis. For the purpose of $(g-2)_\mu$  one needs to
consider only terms like $y_{\mu\ell}\Delta^{\pm\pm}\mu^{\mp} \ell^{\mp}$. 
All other Yukawa couplings are taken to be zero and further, we 
assume $y_{\mu\ell}=y_{\ell\mu}$ as well as $y_{\mu e}=y_{\mu\tau}$. 
One must remain careful while interpreting these $y_{\mu\ell}$ where information 
about the specific model (see Refs.~\cite{delAguila:2013yaa,delAguila:2013mia})
are also embedded.
Our simplified choice of $y_{\mu\ell}$ leaves us with only three free parameters,
namely $y_{\mu\mu},\,y_{\mu e}(=y_{\mu\tau})$ and $\mdd$ relevant for our analysis.
It is apparent from our choice of Yukawa couplings that, along with  $(g-2)_\mu$, 
a few CLFV processes like $\mu \to e\gamma$, $\tau\to \mu \gamma$, $\mu N\to e N^*$ 
($\mu$ to $e$ conversion in atomic nuclei) etc. are automatically switched on. 
We will elaborate this issue further in Sec. \ref{clfv}. 
Finally, we need to write down the relevant terms which lead
to $pp\to\dc\Delta^{\mp\mp}$ process at the LHC. 
The necessary {\it trilinear} couplings between $\dc$ and the electroweak gauge-bosons, in the 
absence of VEV for the new multiplet, are given as~\cite{delAguila:2013mia}:
$2i g_2 \sin \theta_W  \Delta^{\pm \pm} \Delta^{\mp \mp} A_\sigma p^\sigma$ and 
$i(g_2/\cos \theta_W)  (2-{Y} -2 \sin^2 \theta_W) \Delta^{\pm \pm} 
\Delta^{\mp \mp} Z_\sigma p^\sigma$, where $p^\sigma$ is the momentum transfer 
at these vertices. Here, $Y$ is hypercharge of the scalar multiplet that contains $\dc$,
$g_2$ is the $SU(2)$ gauge coupling and $\theta_W$ is Weinberg angle \cite{Agashe:2014kda}.
It is important to note the structure of $\dc\Delta^{\mp\mp}Z_\mu$ vertex, where some knowledge
of the underlying multiplet appears necessary through the hypercharge quantum 
number\footnote{A complete knowledge of the underlying multiplet would also
require information of the isospin.}.

\subsection*{\boldmath Massive neutrinos with a $\dc$}\label{massnu}

In this subsection, as mentioned in the beginning, 
we aim to present a brief discussion about the neutrino mass generation in the 
presence of a $\dc$. Accommodating tiny neutrino masses 
in a model with $\dc$ can be achieved in different ways, e.g.,
in the tree-level from a Type-II seesaw mechanism using an $SU(2)$ 
triplet \cite{Konetschny:1977bn,Senjanovic:1978ev,
Magg:1980ut,Schechter:1980gr,Cheng:1980qt} or in the loop-level
with an $SU(2)$ singlet doubly and another $SU(2)$ singlet singly
$(S^+)$ charged scalar \cite{Babu:1988ki} etc.
However, a $\dc$ gets directly involved in the mechanism of
neutrino mass generation only for the latter model
and thus, we restrict our discussion only for this framework.

It turns out for the aforesaid scenario, better known as the Babu-Zee model \cite{Zee:1980ai,
Zee:1985rj,Zee:1985id,Babu:1988ki}, one needs a $S^+$ along with a $\dc$ to generate
neutrino masses ($m_\nu$) in the two-loop level. Now, let us assume that $y^s_{\ell\ell'},\,y^d_{\ell\ell'}$
represent the {\it{generic real}} Yukawa couplings between the charged leptons and
$S^+,\,\dc$, respectively with the following property: 
$y^{s(d)}_{\ell\ell'}=y^{s(d)}_{\ell'\ell}$.
At this point if we set the scale of new physics (i.e., masses of $S^\pm,\,\dc$
($m_{S^\pm},\,\mdd$) and 
any other relevant parameter) $\sim \mathcal{O}(1$ TeV), just for an example,
then following Ref.~\cite{Babu:1988ki} one can extract the following conditions:

\vspace*{0.1cm} 
\noindent
(i) $\sum y^s_{\mu\ell}y^s_{\ell\mu} 
+ \sum y^d_{\mu\ell}y^d_{\ell\mu} \sim \mathcal{O}(10)$ (from $(g-2)_\mu$~\cite{NYFFELER:2014pta}),

\vspace*{0.1cm}
\noindent
(ii)  $ \sum y^s_{e \ell}y^s_{\ell\mu} + \sum  y^d_{e\ell}y^d_{\ell\mu} 
\lsim \mathcal{O}(0.001)$ (from $\mu\to e \gamma$~\cite{Adam:2013mnn}) and

\vspace*{0.1cm}
\noindent 
(iii) $y^d_{\mu\mu}y^{s^2}_{\mu\tau}(y^{s^2}_{e\tau}
+y^{s^2}_{\mu\tau}+y^{s^2}_{e\mu})/(y^{s^2}_{e\tau}+y^{s^2}_{\mu\tau})  
\sim 0.06$ and $(y^{s^2}_{e\tau}+y^{s^2}_{\mu\tau}) y^d_{\tau\tau} 
\sim 2\times 10^{-4}$ (assuming $m_\nu\sim 0.1$ eV \cite{Ade:2015xua}).

At this point, let us assume that only $y^s_{\mu\mu},\,y^d_{\mu\mu}$ 
are $\sim\mathcal{O}(1)$ while other associated $y^s,\,y^d$ values remain
at least $\lsim 0.1$. Now one can easily satisfy
condition (i) with $y^s_{\mu\mu} \sim 3$ and $y^d_{\mu\mu}\sim 1$. 
These chosen values can safely coexist with condition (ii) as long as (at least)
$y^s_{e\mu},\,y^d_{e\mu}$ $\lsim 0.001$ and $y^s_{e\tau}, \,y^d_{e\tau}$ $\lsim 0.01$. 
With these estimations one can rewrite condition (iii), up to a good
approximation, as: $y^d_{\mu\mu}y^{s^2}_{\mu\tau} \sim 0.06$ and 
$y^{s^2}_{\mu\tau} y^d_{\tau\tau} \sim 2\times 10^{-4}$. The former
is trivially satisfied with $y^d_{\mu\mu}\sim 1$ and
$y^s_{\mu\tau}\lsim \mathcal{O}(0.1)$. The latter 
with $y^s_{\mu\tau}\lsim \mathcal{O}(0.1)$
hints $y^d_{\tau\tau}\sim \mathcal{O}(0.01)$. Making the scale
of new physics $\sim\mathcal{O}(500$ GeV) one can get 
similar results, however, with reduced {\it{upper bounds}} on 
the concerned Yukawa couplings.

Collectively, playing with a larger set of Yukawa couplings,
e.g., imposing hierarchical structures between $y^d$ and $y^s$
as well as among different flavour indices, one can always
satisfy all the three aforementioned criteria. It is
evident from conditions (i) - (iii), that the constraints 
from neutrino mass {\it{simultaneously}} put bounds on the off-diagonal $y^s$ and 
diagonal $y^d$. On the contrary, limits from $(g-2)_\mu$ and CLFV processes
constrain {\it{independently}} both the diagonal and off-diagonal $y^s,\,y^d$ couplings. Thus,
without emphasising the neutrino physics, one can safely work in a scenario when $y^s\to 0$.
The remaining Yukawa couplings, namely $y^d$s, however, remain
tightly constrained from the experimental limits on
muon $(g-2)$ and CLFV processes. An analysis of the said kind,
thus, provides maximum estimates for the associated $y^d$-type
couplings. Any further attempt to add additional requirements
for the same model, like neutrino mass generation, would
only provide reduced upper bounds on the involved $y^d$s. Hence, for 
the rest of the work we keep on working with only $y_d$-type 
Yukawa couplings (henceforth read as $y_{\ell\ell'}$), imposing a 
minimal structure essential to account for the muon $(g-2)$ anomaly. 
We note in passing that, unless compensated by the relative mass 
hierarchies, the contribution to muon 
$(g-2)$ from a $\dc$ normally exceeds the same from a $S^\pm$ 
since $\Delta^{\pm\pm}\Delta^{\mp\mp}\gamma$, $S^\pm S^\mp\gamma$
vertices are sensitive to the electric charges of the concerned fields.

\section{\boldmath Muon $(g-2)$ with $\dc$}\label{gm2}
Precision measurements of the different low-energy processes
always provide an acid test for the BSM theories. Observation of
any possible disparity for these processes, compared
to the corresponding SM predictions, provides a golden opportunity to explore as well as
constrain various BSM models. The observed discrepancy in the
anomalous magnetic moment of muon, $\Delta a_\mu$ 
is a very intriguing example of this kind.

\begin{figure}
\begin{center}
\includegraphics[width=8.0cm,height=1.9cm]{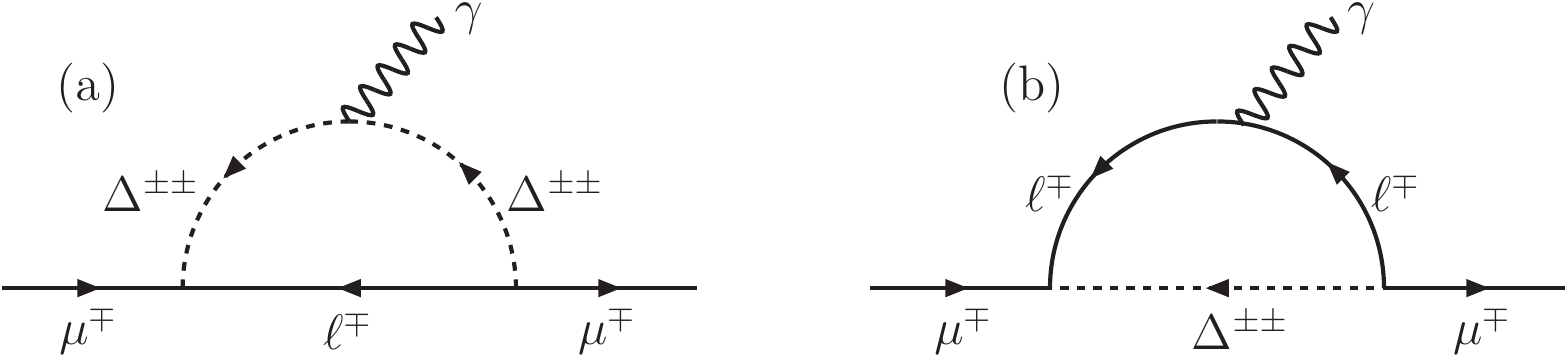}
\caption{\label{fig0}Possible Feynman diagrams showing contributions to the muon anomalous magnetic 
moment through the exchange of a doubly charged scalar and the selected $y_{\mu\ell}$. 
The arrows represent the direction of electric charge flow.}
\end{center}
\end{figure}

In the SM, anomalous magnetic moment of muon is associated with the coupling 
$ig_2 \sin \theta_W$ $\overline{\mu} \gamma^\sigma \mu A_\sigma$.
However, even including the higher order contributions within the SM one can not 
explain the observed discrepancy $\Delta a_\mu = $ $a^{exp}_\mu$ \cite{Bennett:2006fi,
Roberts:2010cj} - $a^{th}_\mu$ \cite{Prades:2009tw,Nyffeler:2010rd,Davier:2010nc,Hagiwara:2011af}.
Here $a^{exp}_\mu$ is the experimentally measured value of $(g-2)_\mu$
and $a^{th}_\mu$ is the theoretical estimate of $(g-2)_\mu$ in the context of the SM.
The latest numerical value\footnote{One should note that depending on the calculation of $a^{th}_\mu$, the
value of $\Delta a_\mu$ may change as pointed out in Ref.~\cite{KNECHT:2014bsa}.} 
following Ref.~\cite{NYFFELER:2014pta} is given by
\begin{equation}\label{gm2value}
\Delta a_\mu = a^{exp}_\mu -a^{th}_\mu = (29.3 \pm  9.0) \times 10^{-10}.
\end{equation} 
The presence of any possible BSM contribution will generally affect
the $\mu \to \mu \gamma$ process at the loop level. In the presence
of a $\dc$, the new BSM contributions to $(g-2)_\mu$ are shown in Fig.~\ref{fig0}.
%
%
From this figure one can see that a $\dc$ can contribute to
$(g-2)_\mu$ through two possible ways. Following Refs.~\cite{Leveille:1977rc,Moore:1984eg}, 
 new contribution\footnote{
We derive contributions from a $\dc$ (see appendix) in a way 
similar to the {\it Higgs type contribution}
as shown in these reference.} to $(g-2)_\mu$, in the presence of $\dc$ and 
allowing all the charged leptons in the loop, is given by:
\begin{eqnarray} \label{mugm2full}
&&\Delta a_\mu =   \frac{f_m m_\mu^2 y_{\mu \ell}^2}{8\pi^2}\nonumber \\
&&\times\Bigg[   \int_{0}^{1} 
d\rho\,\,\frac{ 2 (\rho +\frac{m_\ell}{m_\mu})(\rho^2-\rho)}{[m_\mu^2 \rho^2 
+(m_{\Delta^{\pm\pm}}^2-m_\mu^2)\rho+(1-\rho)m_{\ell}^2]}   \nonumber \\
& & - \int_{0}^{1} d\rho\,\,\frac{  (\rho^2-\rho^3 +\frac{m_\ell}{m_\mu} \rho^2)}{[m_\mu^2 \rho^2 
+(m_{\ell}^2-m_\mu^2)\rho+(1-\rho)m_{\Delta^{\pm\pm}}^2]}\Bigg],
\end{eqnarray}
where $m_\mu$ is the mass of muon \cite{Agashe:2014kda} and $m_{\ell}$ is 
the mass of charged lepton $``\ell$''. We have also assumed 
real $y_{\mu\ell}$, so that $|y_{\mu\ell}|^2=y_{\mu \ell}^2$. Further
details of the computation are relegated to the appendix. In Eq.~(\ref{mugm2full}),
$f_{m}$ is equal to $1$ for $\ell=e,\,\tau$ while equals to $4$ for $\ell=\mu$. 
This multiplicative factor appears due to the presence of two identical 
fields in the interaction term.

\section{CLFV and \boldmath $\dc$}\label{clfv}
%
The most general Yukawa interactions between the charged leptons
and a $\dc$ contains off-diagonal Yukawa couplings  that are instrumental
in producing CLFV processes like $\ell_i\to \ell_j\gamma$, 
$\ell_a\to\ell_b\ell_c\ell_d$ etc. In this article, however,
we have assumed a minimal set of Yukawa couplings focusing on 
the muon anomalous magnetic moment. Thus, as already stated
in Sec. \ref{Scenario}, our Yukawa sector contains only
$y_{\mu\mu},\,y_{\mu e}$ and $y_{\mu\tau}$, with $y_{\mu e}=y_{\mu\tau}$.
Such a parameter choice would allow {\it only six} CLFV processes, 
namely $\mu \to e \gamma,\,\tau \to e \gamma,\,
\tau \to \mu \gamma,\,\tau \to 3\mu,\,\tau \to e \mu^+ \mu^-$ 
and $\mu N \to e N^{*}$ at the {\it respective leading orders}, 
that too with {\it only a few possible} diagrams.
For the clarity of reading, we describe all such diagrams in Fig.~\ref{fig1n}.
At this stage, it appears crucial to explain the phrase ``only six CLFV processes 
at the leading  orders'' in order to ameliorate any possible delusion. It is absolutely
true that the chosen set of $y_{\mu\ell}$ forbids tree-level processes
like $\mu\to 3e$, $\tau\to \mu e^+e^-$ through an off-shell $\dc$, as 
sketched for $\tau\to e \mu^+\mu^-$ process in diagram (e) of Fig.~\ref{fig1n}\footnote{
These processes can show-up at the one-loop level 
via $Z/\gamma^*$ mediator. Depending on the set of involved parameters
process like $\ell_a\to \ell_i \ell_j\ell_k$ and also $\mu$ - $e$ conversion
in the nuclei may enjoy an extra enhancement from $Z$-penguin \cite{Hirsch:2012ax}. The latter
can offer severe constraints on the parameter space compared to
$\ell_i\to \ell_j \gamma$ processes \cite{Wilczek:1977wb,Treiman:1977dj,
Altarelli:1977zq,Marciano:1977cj,Raidal:1997hq}, which normally holds true in
the reverse order. However, following Ref.~\cite{Raidal:1997hq}
one can conclude that such enhancement will not modify the scale
of new physics ($\mdd$ in our analysis) by orders of magnitudes.
We, thus, do not consider these ``enhancements'' in our present analysis.}.
All the relevant branching fractions $(Br)$ for the set of 
processes shown in Fig.~\ref{fig1n} are given  
below \cite{Chun:2003ej,Kakizaki:2003jk,Akeroyd:2009nu,Chakrabortty:2012vp}:

\begin{figure}
\begin{center}
\includegraphics[width=7.2cm,height=1.85cm]{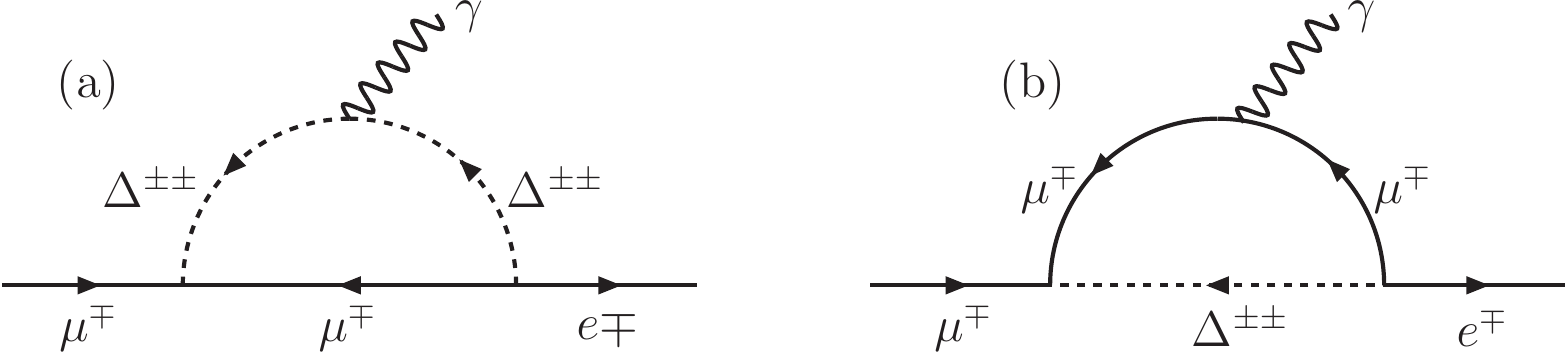}
\vspace*{0.2cm}
\includegraphics[width=7.2cm,height=1.85cm]{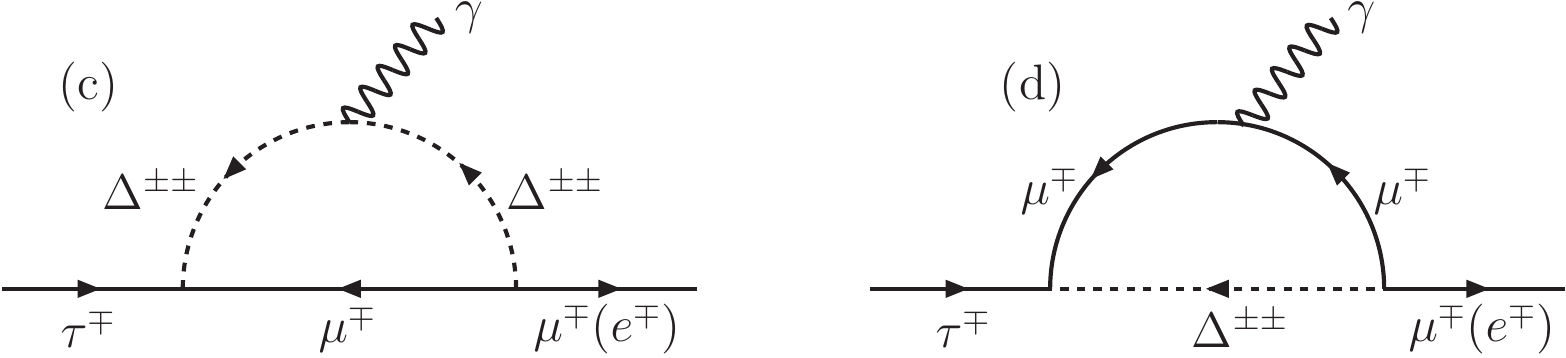}
\vspace*{0.2cm}
\includegraphics[width=3.5cm,height=1.0cm]{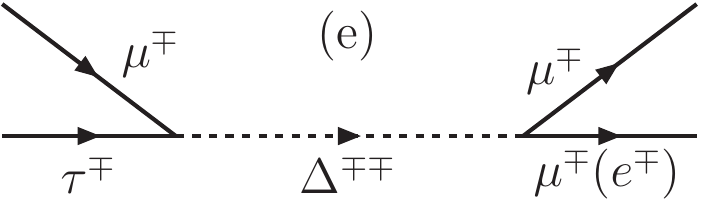}
\vspace*{0.2cm}
\includegraphics[width=7.2cm,height=1.85cm]{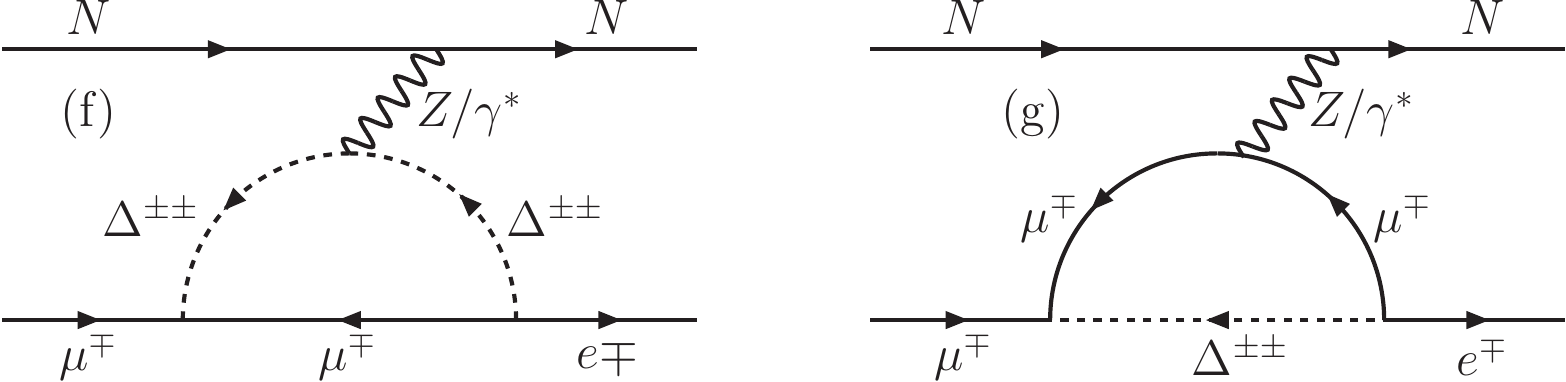}
\caption{\label{fig1n}Possible Feynman diagrams showing contributions to 
$\mu \to e \gamma\, ({\rm a,\,b}), \tau \to e \gamma, \tau \to \mu \gamma\, ({\rm c,\,d}), \tau \to 3\mu, 
\tau \to e \mu^+ \mu^-\, ({\rm e})$ and $\mu N \to e N^{*}\, ({\rm f,\,g})$ processes in the presence of a 
$\dc$ and the selected $y_{\mu\ell}$s. Here, $N$ represents the concerned atomic nucleus. 
The arrows represent the direction of electric charge flow.}
\end{center}
\end{figure}

\begin{eqnarray}\label{clfv1}
& &Br(\mu \to e \gamma) = \frac{27 \alpha_{em}|(yy^\dagger)_{e\mu}|^2}{64\pi G_F^2 m_{\Delta^{\pm \pm}}^4} 
~ ~Br(\mu \to e \bar{\nu}_e \nu_\mu ),\\
& &Br(\tau \to e \gamma) = \frac{27 \alpha_{em}|(yy^\dagger)_{e\tau}|^2}{64\pi G_F^2 m_{\Delta^{\pm \pm}}^4} 
~Br(\tau \to e \bar{\nu}_e \nu_\tau ),\\
& &Br(\tau \to \mu \gamma) = \frac{27 \alpha_{em}|(yy^\dagger)_{\mu \tau}|^2}{64\pi G_F^2 m_{\Delta^{\pm \pm}}^4} 
~Br(\tau \to \mu \bar{\nu}_\mu \nu_\tau ),\\
& &Br(\tau \to 3\mu) = \frac{|y_{\tau \mu }|^2 ~|y_{\mu \mu }|^2}{4 G_F^2 m_{\Delta^{\pm \pm}}^4} 
~~Br(\tau \to \mu \bar{\nu}_\mu \nu_\tau ),~{\rm and}\\
& &Br(\tau \to \mu \mu e) = \frac{|y_{\tau \mu }|^2 ~|y_{\mu e }|^2}{4 G_F^2 m_{\Delta^{\pm \pm}}^4} 
~~Br(\tau \to \mu \bar{\nu}_\mu \nu_\tau ),
\end{eqnarray}
where $\alpha_{em}=g^2_2 \sin^2\theta_W/4\pi$, $G_F$ is the Fermi constant \cite{Agashe:2014kda},
$Br(\mu \to e \bar{\nu}_e \nu_\mu )=100\%$, $Br(\tau \to e \bar{\nu}_e \nu_\tau)=17.83\%$
and $Br(\tau \to \mu \bar{\nu}_\mu \nu_\tau)=17.41\%$ \cite{Agashe:2014kda}.
 
The rate of $\mu\to e$ conversion in atomic nuclei with the chosen set of $y_{\mu\ell}$  
is written as
 
\bea\label{mu2e}
R(\mu N \to && e N^{*})=\frac{(\alpha_{em} m_{\mu})^5 Z_{\rm eff}^4 Z |F(q)|^2}
{4\pi^4 m_{\Delta^{\pm \pm}}^4 \Gamma_{\rm capt}}\nonumber\\
&& \times \Big|\frac{y^\dagger_{e\mu} y_{\mu\mu} F(r,s_\mu)}{3}  
-\frac{3(y^\dagger y)_{e\mu}}{8}  \Big|^2,{\rm where~}
\eea
%
\bea\label{mu2e-parts}
F(r,s_\mu)&&={\rm ln}~s_\mu + \frac{4s_\mu}{r}+(1-\frac{2 s_\mu}{r})\nonumber\\
&& \times \sqrt{(1+\frac{4s_\mu}{r})} ~~
{\rm ln} \frac{\sqrt{(1+\frac{4s_\mu}{r})} + 1}{\sqrt{(1+\frac{4s_\mu}{r})} -1},\nonumber\\
&& r=-\frac{q^2}{m_{\Delta^{\pm \pm}}^2}, 
s_\mu = \frac{m_\mu^2}{m_{\Delta^{\pm \pm}}^2}.
\eea
Here, $Z$ is the atomic number of the concerned nucleus. Values 
of $Z_{\rm eff}$, $\Gamma_{\rm capt}$ and $F(q^2\backsimeq-m^2_{\mu})$
for the  different atomic nuclei can be obtained from Ref.~\cite{Kitano:2002mt}.

Finally, before we start discussing our results in the next section,
we summarise the present and the expected future limits of the considered
CLFV processes in Table \ref{lfv-limits}.
%

\begin{table}
\begin{center}
\scriptsize
\begin{tabular}{ c c c} \hline\hline
 CLFV processes & Present Limit & Future Limit \\ 
\hline 
$\rm{BR}(\mu\rightarrow e\gamma)$ & $5.7\times 10^{-13}$ \cite{Adam:2013mnn} 
& $6.0\times 10^{-14}$ \cite{Baldini:2013ke} \\ 
$\rm{BR}(\tau\rightarrow e\gamma)$ & $3.3\times 10^{-8}$ \cite{Aubert:2009ag} 
& $3.0\times 10^{-9}$ \cite{Aushev:2010bq}   \\
$\rm{BR}(\tau\rightarrow\mu\gamma)$ & $4.4\times 10^{-8}$ \cite{Aubert:2009ag} 
& $3.0\times 10^{-9}$ \cite{Aushev:2010bq}   \\
$\rm{BR}(\tau\rightarrow 3\mu)$ & $2.1\times 10^{-8}$ \cite{Hayasaka:2010np} 
& $1.0\times 10^{-9}$ \cite{Aushev:2010bq}   \\
$\rm{BR}(\tau\rightarrow e \mu^+\mu^-)$ & $2.7\times 10^{-8}$ \cite{Hayasaka:2010np} 
& $1.0\times 10^{-9}$  \cite{Aushev:2010bq}  \\
R($\mu N\rightarrow eN^*$)  & $7.0\times 10^{-13}$  \cite{Bertl:2006up} 
& $2.87\times 10^{-17}$ \cite{Bartoszek:2014mya}  \\
(for ${\rm Au}$) & &   \\
\hline \hline
\end{tabular}
\caption{\label{lfv-limits}The present and the expected future limits of the concerned CLFV processes.}
\end{center}
\end{table} 

\section{Results}\label{result}
\begin{center}
\begin{figure*}[htb]
\includegraphics[width=6.25cm]{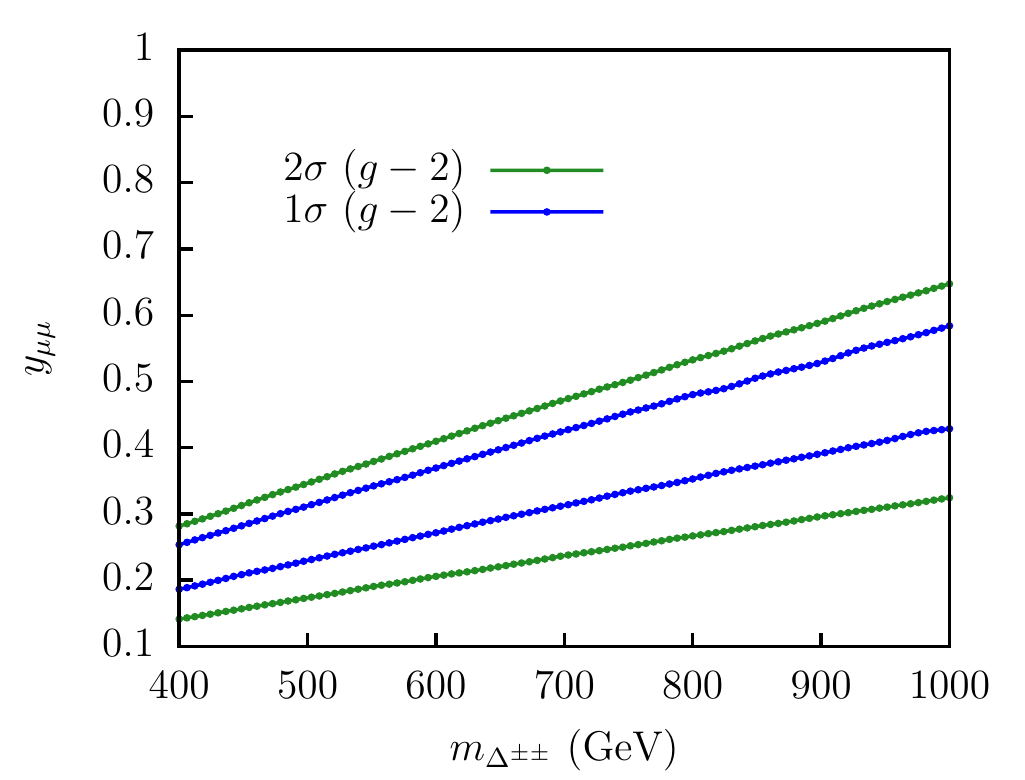}
\includegraphics[width=6.25cm]{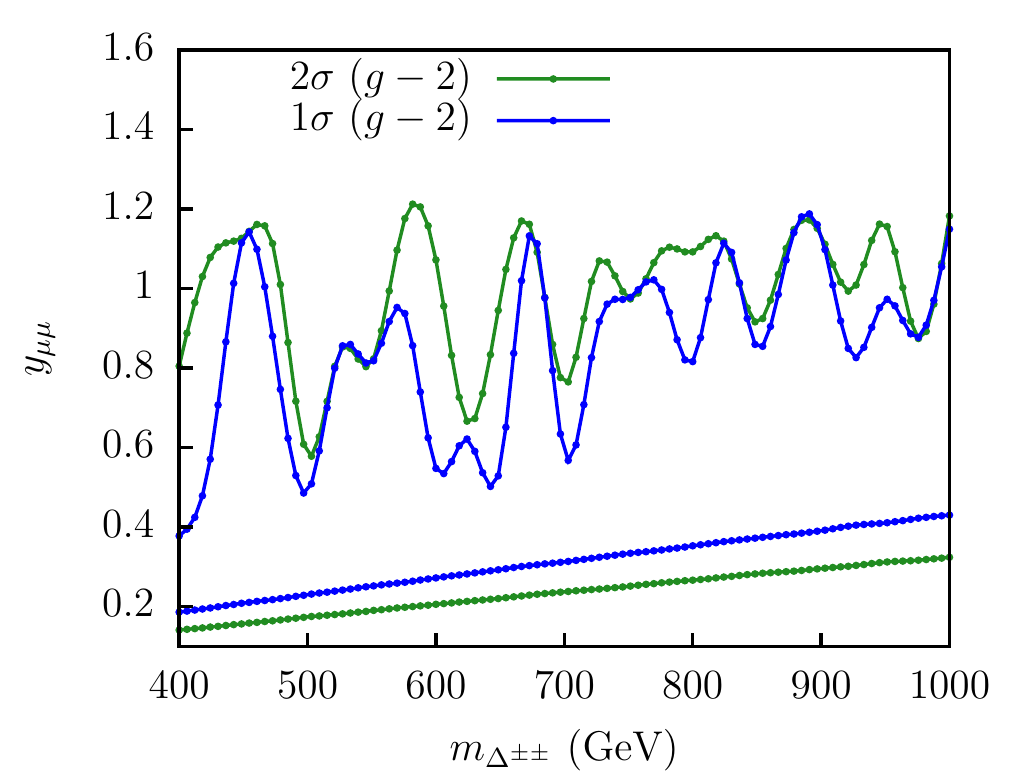}
\caption{\label{fig1}Correlations between $\mdd$ and $y_{\mu\mu}$ 
for the allowed one and two sigma ranges of $\Delta a_\mu$. In the left plot
$\Delta a_\mu$ is originating solely from $y_{\mu\mu}$
while in the right plot contributions from other off-diagonal $y_{\mu\ell}$s
are also included. The blue and the green lines
represent the one and two sigma bands of the allowed
$\Delta a_\mu$ values, as given by Eq.~(\ref{gm2value}).}
\end{figure*}
\end{center}

We initiate exploring our findings with the muon anomalous magnetic moment in
the context of BSM input parameters $\mdd$ and $y_{\mu\ell}~(\equiv y_{\ell\mu})$. 
A self-developed \texttt{FORTRAN} code has been used for the purpose of numerical 
analyses. In our investigation we perform a scan over three free parameters
$y_{\mu\mu}$, $y_{\mu e} (\equiv y_{\mu\tau})$ and $\mdd$ in the following ranges:
$10^{-4}\lsim y_{\mu\mu}, y_{\mu e}\lsim 1.2$ and 
$400~{\rm GeV}$ $\lsim \mdd \lsim 1000~{\rm GeV}$, respectively.
For the analysis of $(g-2)_\mu$ we do not consider any constraints from
the list of CLFV processes shown in Table \ref{lfv-limits}.
In Fig. \ref{fig1},
we plot the variation of $y_{\mu\mu}$ with $\mdd$ when (i) only $y_{\mu\mu}$
is contributing to $\Delta a_\mu$ (left plot), that is $\ell$=$\mu$ in Fig.~\ref{fig0}
and, (ii) all the chosen $y_{\mu\ell}$s are contributing to $\Delta a_\mu$ (right plot).
The left plot of Fig.~\ref{fig1} shows a copacetic correlation between $\mdd$
and $y_{\mu\mu}$, as expected for an analysis with only two free parameters [see Eq.~(\ref{mugm2full})
with $y_{\mu\ell}=0$ for $\ell\neq\mu$]. The smooth increase of $y_{\mu\mu}$
with larger $\mdd$ values is also well understood from the same equation since $y_{\mu\mu}$ appears
in the numerator while $\mdd$ in the denominator. Hence, larger $y_{\mu\mu}$
values appear a must to satisfy the constraint on $\Delta a_\mu$ with increasing $\mdd$.
The blue and the green lines represent
lower and upper bounds of the allowed one and two sigma (1$\sigma$-2$\sigma$) ranges for $\Delta a_{\mu}$ 
[see Eq.~(\ref{gm2value})], respectively. From the left plot one can also
extract the possible range for $y_{\mu\mu}$, i.e., between $0.1-0.65$ when
$\mdd$ varies within $400~{\rm GeV}-1000$ GeV. The astonishing correlation between
$y_{\mu\mu}$ and $\mdd$ gets distorted when one switches on the other 
off-diagonal Yukawa couplings, namely $y_{\mu e}$ and $y_{\mu\tau}$, as
shown in the right plot of Fig.~\ref{fig1}. These distortions are
apparent only for the upper bands of allowed one and two sigma $\Delta a_\mu$ values
while the lower bands remain practically the same as the scenario with only $y_{\mu\mu}$.
Two conclusions become apparent from the right plot of Fig.~\ref{fig1}:
(1) off-diagonal Yukawa couplings can produce significant contributions to $\Delta a_\mu$
and, (2) these new contributions are normally negative and thus, one needs larger
$y_{\mu\mu}$ values to accommodate the $\Delta a_\mu$ data.
At the same time,  the similarity of the lower one and two sigma lines,
in both of the plots, implies that contributions from the off-diagonal $y_{\mu\ell}$s are 
typically smaller compared to the same from $y_{\mu\mu}$.
Unlike the left plot, here one does not get a smooth increase in $y_{\mu\mu}$
value with increasing $\mdd$. One can however, still estimate a range
for $y_{\mu\mu}$, i.e., $0.1-1.2$  for $400~{\rm GeV}\leq\mdd\leq 1000~{\rm GeV}$.

\begin{center}
\begin{figure*}
\includegraphics[width=6.25cm,angle=0]{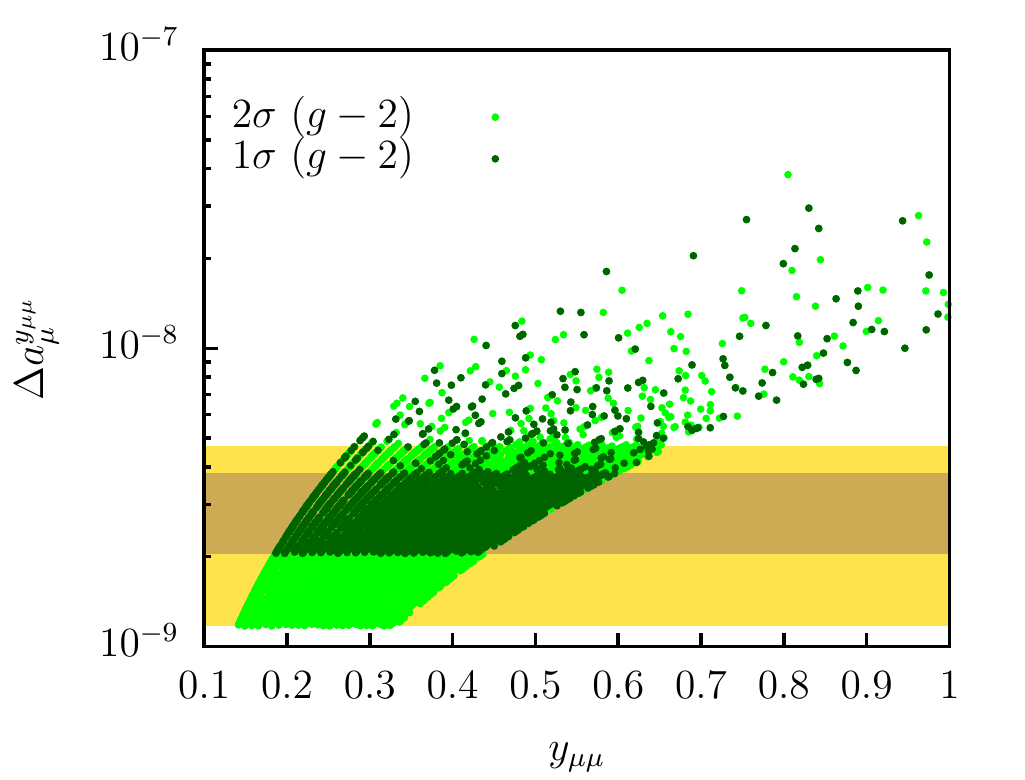}
\includegraphics[width=6.25cm,angle=0]{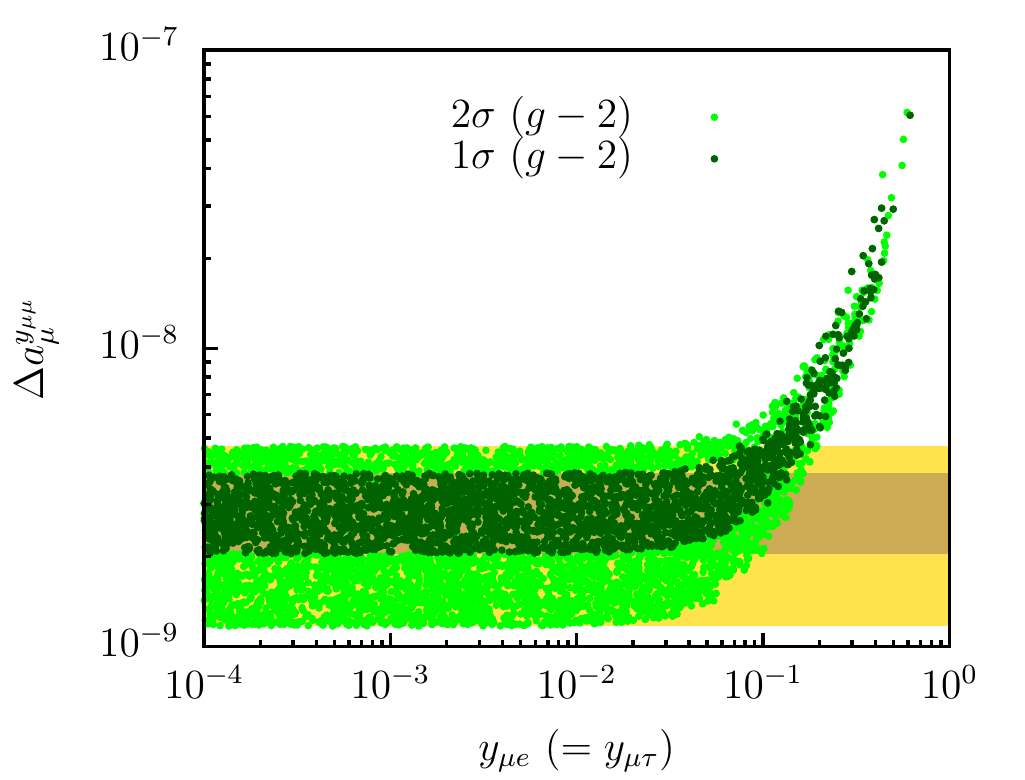}\\
\includegraphics[width=6.25cm,angle=0]{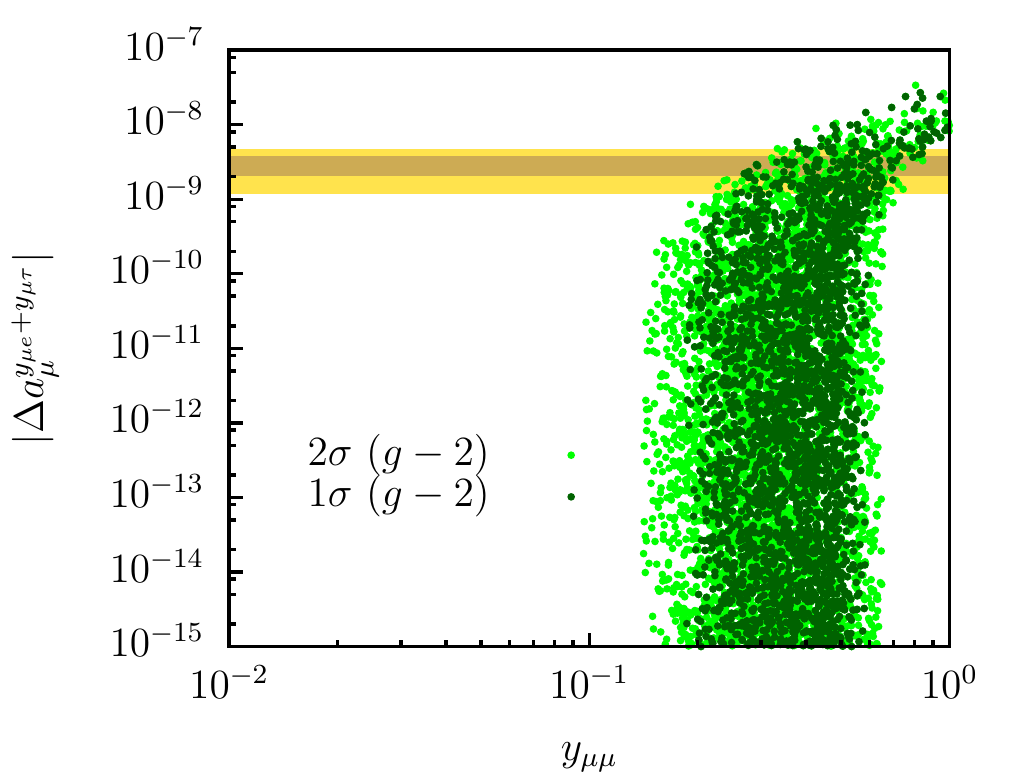}
\includegraphics[width=6.25cm,angle=0]{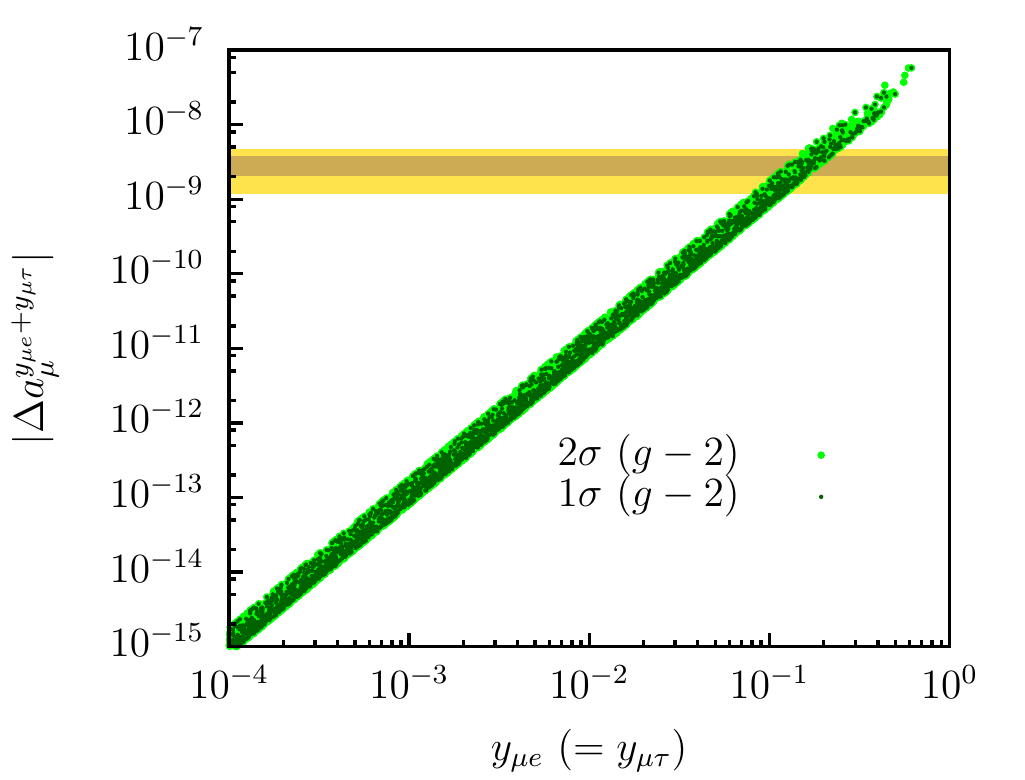}
\caption{\label{fig2} Plots showing variations of $\Delta a^{y_{\mu\mu}}_\mu$
with $y_{\mu\mu}$ (top-left), $y_{\mu e}(\equiv y_{\mu\tau})$ (top-right) and the changes 
of $|\Delta a^{y_{\mu e}+y_{\mu\tau}}_\mu|$
with $y_{\mu\mu}$ (bottom-left), $y_{\mu e} (\equiv y_{\mu\tau})$ (bottom-right).
The quantities $\Delta a^{y_{\mu\mu}}_\mu$ and $|\Delta a^{y_{\mu e}+y_{\mu\tau}}_\mu|$
are explained in the text. The light-brown and golden coloured bands represent
the measured $1\sigma$ and $2\sigma$ ranges of $\Delta a_\mu$ (see Eq.~(\ref{gm2value})).
The deep-green (light-green) coloured point represents whether it satisfies the constraint on
 $\Delta a_\mu$ at the $1\sigma~(2\sigma)$ interval.
We consider $400~{\rm GeV}\lsim \mdd \lsim 1000~{\rm GeV}$ for these plots.}     
\end{figure*}
\end{center}

In order to understand the relative contributions from $y_{\mu\mu}$
and $y_{\mu e},\,y_{\mu\tau}$ to the computation of $\Delta a_\mu$,
we  plot the four possible variations in Fig.~\ref{fig2}.
We consider the same range of $\mdd$, i.e., 
$400~{\rm GeV}\lsim\mdd\lsim1000~{\rm GeV}$ for these plots, similar to Fig.~\ref{fig1}.
All  these data points (deep and light greens) satisfy the one and two sigma bounds
on $\Delta a_\mu$ (represented by the light-brown and golden coloured
bands, respectively), as given in Eq.~(\ref{gm2value}). Two plots 
in the top row of Fig.~\ref{fig2} show the variations of
$\Delta a^{y_{\mu\mu}}_\mu$ with respect to $y_{\mu\mu}$ and the off-diagonal
Yukawa couplings. Two of the bottom row plots represent the same
but for $|\Delta a^{y_{\mu e}+y_{\mu \tau}}_\mu|$.
Here, $\Delta a^{y_{\mu\mu}}_\mu$ is that part of $\Delta a_\mu$
which arises solely from $y_{\mu\mu}$ while $|\Delta a^{y_{\mu e}+y_{\mu\tau}}_\mu|$
represents the same from $y_{\mu\ell}$ with $\ell\neq \mu$ [see Eq.~(\ref{mugm2full})].
From the top-left plot of Fig.~\ref{fig2}, it is evident that
in the presence of off-diagonal Yukawas, $y_{\mu\mu}\gsim 0.3$
can yield a large contribution to muon $(g-2)$ beyond $2\sigma$. Thus, if we assume 
that contribution to $\Delta a_\mu$ arises solely from $y_{\mu\mu}$, 
i.e., $\Delta a_\mu \approx \Delta a^{y_{\mu\mu}}_\mu $, all points above
the golden band remain experimentally excluded. The situation remains
the same for $\Delta a^{y_{\mu\mu}}_\mu$ in
the context of off-diagonal $y_{\mu\ell}$ when $y_{\mu e}$ or 
$y_{\mu \tau}\gsim 0.2$ (top-right plot for Fig.~\ref{fig2}).
Beyond $y_{\mu e}=0.2$, the sizeable but opposite $sign$ contributions
(-$|\Delta a^{y_{\mu e}+y_{\mu\tau}}_\mu|$) from $y_{\mu\ell}$ 
adjust the positive over-growth of $\Delta a^{y_{\mu\mu}}_\mu$ 
beyond $2\sigma$ for $y_{\mu\mu}\gsim 0.3$, as shown in the bottom-right plot of Fig.~\ref{fig2}.
One more observation is apparent from the bottom-right plot of Fig.~\ref{fig2},
that  the contribution from $|\Delta a^{y_{\mu e}+y_{\mu\tau}}_\mu|$
in the determination of $\Delta a_\mu$ is practically negligible for $y_{\mu e}\lsim 0.01$. 
On the contrary, as can be seen from the bottom-left plot of Fig.~\ref{fig2},
that $|\Delta a^{y_{\mu e}+y_{\mu\tau}}_\mu|$ shows
hardly any sensitivity to $y_{\mu\mu}$ below $y_{\mu\mu}\lsim 0.3$. Only
in the regime of $y_{\mu\mu}\gsim 0.3$, $|\Delta a^{y_{\mu e}+y_{\mu\tau}}_\mu|$
grows with $y_{\mu\mu}$. This growth becomes prominent for $y_{\mu\mu}\gsim 0.7$.
So one can conclude that:

\begin{list}{}{}

 \item 
 (1) For the region $y_{\mu\mu}\lsim 0.3$, $y_{\mu e} (\equiv y_{\mu\tau})\lsim 0.01$,
$\Delta a_\mu = \Delta a^{y_{\mu\mu}}_\mu + \Delta a^{y_{\mu e}+y_{\mu\tau}}_\mu
\approx \Delta a^{y_{\mu\mu}}_\mu$. This is the reason why the lower
one and two sigma lines for the two plots of Fig.~\ref{fig1} remain
almost unaltered.

\item
(2) In a tiny region: $0.3\lsim y_{\mu\mu}\lsim 0.7$,
$0.01\lsim y_{\mu e} (\equiv y_{\mu\tau})\lsim 0.2$, both of the contributions
remain comparable to the measured $\Delta a_\mu$ [see Eq.~(\ref{gm2value})],
i.e., $|\Delta a^{y_{\mu e}+y_{\mu\tau}}_\mu|\sim \Delta a^{y_{\mu\mu}}_\mu$
$\sim {\cal{O}}(\Delta a_\mu)$. Hence, the measured constraint on $\Delta a_\mu$
appears feasible after a {\it tuned} cancellation between
$\Delta a^{y_{\mu\mu}}_\mu$ and $\Delta a^{y_{\mu e}+y_{\mu\tau}}_\mu$.

\item
(3) Finally, in the region with $y_{\mu\mu}\gsim 0.7$,
$y_{\mu e} (\equiv y_{\mu\tau})\gsim 0.2$, both of the contributions
are larger than the measured $\Delta a_\mu$ (beyond
the golden band at $2\sigma$ level). In other words,
$|\Delta a^{y_{\mu e}+y_{\mu\tau}}_\mu|\sim \Delta a^{y_{\mu\mu}}_\mu$
$\gg {\cal{O}}(\Delta a_\mu)$. Clearly, for this region,
the parameter space that remains compatible with the measured constraint of $\Delta a_\mu$
appears through a {\it much-tuned} cancellation between
$\Delta a^{y_{\mu\mu}}_\mu$ and $\Delta a^{y_{\mu e}+y_{\mu\tau}}_\mu$.
 
\end{list}

These last two features are also reflected in the erratic
variation of $y_{\mu\mu}$, as shown in the right panel of Fig.~\ref{fig1}.

\begin{figure}[ht]
\begin{center}
\includegraphics[width=6.25cm,angle=0]{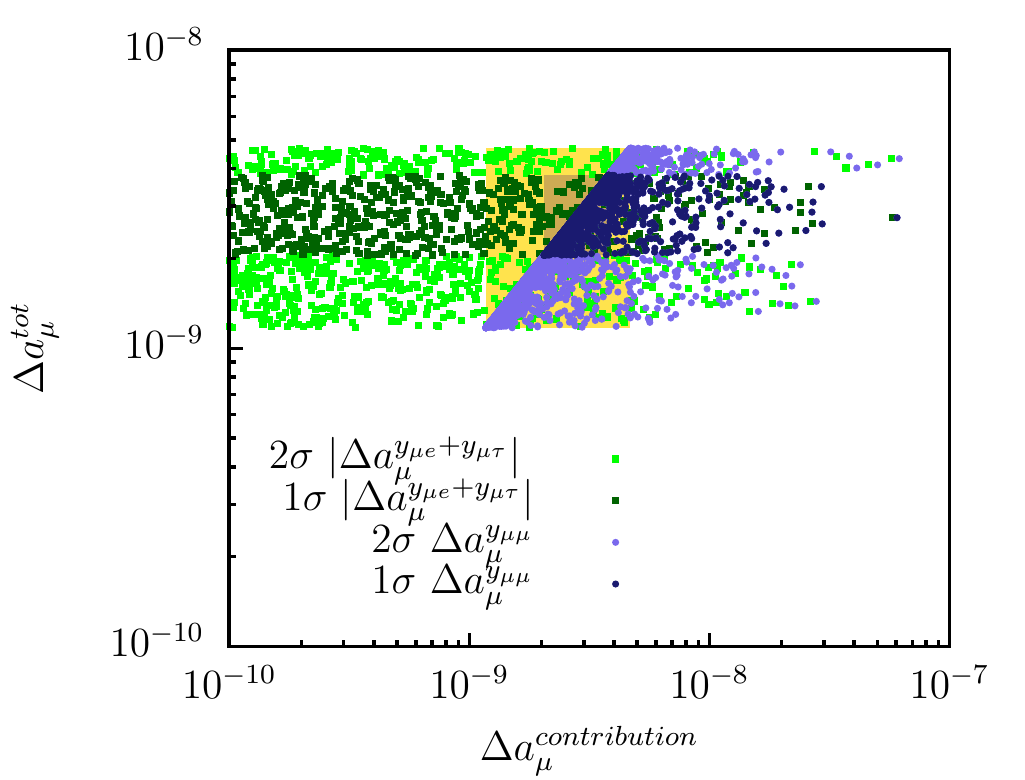}
\caption{\label{fig4} Plot showing variations
of the individual components, i.e., $|\Delta a^{y_{\mu e}+y_{\mu\tau}}_\mu|$
and $\Delta a^{y_{\mu\mu}}_\mu$ with the total $\Delta a_\mu (\equiv \Delta a^{tot}_\mu)$. 
The deep-green (light-green) coloured point represents whether it satisfies the constraint of 
$\Delta a_\mu$ at the $1\sigma~(2\sigma)$ interval for $|\Delta a^{y_{\mu e}+y_{\mu\tau}}_\mu|$.
The deep-blue (light-blue) coloured points represent the same
 for $\Delta a^{y_{\mu\mu}}_\mu$. Remaining details are the same as Fig.~\ref{fig2}.} 
\end{center}
\end{figure}

A pictorial representation of these three aforesaid observations
is shown in Fig.~\ref{fig4}. Here, we plot the variations
of individual components, i.e., $|\Delta a^{y_{\mu e}+y_{\mu\tau}}_\mu|$
and $\Delta a^{y_{\mu\mu}}_\mu$ with the total $\Delta a_\mu (\equiv \Delta a^{tot}_\mu)$.  
It is apparent from this plot that the contribution of $\Delta a^{y_{\mu\mu}}_\mu$
in the evaluation of $\Delta a^{tot}_\mu$ is either the leading one 
(regime of overlap with the golden coloured band at the $2\sigma$ interval)
or overshooting. On the other hand, for a novel region of the 
parameter space $|\Delta a^{y_{\mu e}+y_{\mu\tau}}_\mu|$ remains subleading 
(left-hand side of the golden coloured band) or comparable to $\Delta a^{y_{\mu\mu}}_\mu$ 
(regime of overlap with the golden band at the $2\sigma$ interval). Further, for 
$y_{\mu\ell}(\ell \neq \mu)\gsim 0.2$, $|\Delta a^{y_{\mu e}+y_{\mu\tau}}_\mu|$ 
can also overshoot $\Delta a^{tot}_\mu$ like $\Delta 
a^{y_{\mu\mu}}_\mu$ (right-hand side of the golden band). However, this excess is 
opposite in sign to that of the $\Delta a^{y_{\mu\mu}}_\mu$ and thus, together they respect
the $2\sigma$  constraint on $(g-2)_\mu$.

The discussion presented so far in the context of $\Delta a_\mu$,
using the information available from Fig.~\ref{fig1}, Fig.~\ref{fig2}
and Fig.~\ref{fig4}, can be summarised as follows:

\begin{list}{}{}
 
\item
(1)  For most of the parameter space, the dominant contribution to $\Delta a_\mu$ 
is coming from $y_{\mu\mu}$, irrespective of $y_{\mu\ell}~(\ell\neq \mu)$ or $\mdd$.
This region is $y_{\mu e} (\equiv y_{\mu \tau})\lsim 0.01$, $0.1\lsim y_{\mu\mu}\lsim 0.3$.

\item
(2)  The contribution of $y_{\mu\ell}$ in $\Delta a_\mu$ is always
negative and practically negligible till $y_{\mu\ell}\sim 0.01$.
In the range of $0.01\lsim y_{\mu\ell} \lsim 0.2$, $y_{\mu\ell}$
can yield a contribution to $(g-2)_\mu$ comparable to that from $y_{\mu\mu}$
(i.e., when $0.3\lsim y_{\mu\mu}\lsim 0.7$)
but with an opposite sign. Lastly, beyond $y_{\mu\ell}\sim 0.2$,
a large negative contribution from this parameter helps to nullify
the positive overshooting contribution to $\Delta a_\mu$ from $y_{\mu\mu}$
with $y_{\mu\mu}>0.7$.

\item 
(3)  Depending on the chosen range of $\mdd$, 
i.e., $400~{\rm GeV}\lsim \mdd \lsim 1000~{\rm GeV}$, one can extract the upper
bounds for the parameters $y_{\mu\mu}$ and $y_{\mu\ell} (\ell \neq \mu)$
from our analyses as $1.2$ [see right-panel plot of Fig.~\ref{fig1}] 
and $0.6$ (see top-right plot of Fig.~\ref{fig2}), respectively.
These are the absolute possible upper limits of the respective parameters,
as extracted through a simplified  analysis. Adding other
off-diagonal Yukawa couplings or introducing complex
phases will in general result smaller upper bounds for the concerned
parameters. The only trivial way to raise\footnote{In the same
spirit one can consider a lower $\mdd$ to reduce the upper bounds
on $y_{\mu\mu},\,y_{\mu\ell}$. However, $\mdd$ below $400$ GeV
is already at the edge of the experimentally excluded regions \cite{ATLAS:2014kca}.} these bounds
is to consider a higher $\mdd$. This in turn would yield a smaller
production cross-section for the process $pp\to \dc\Delta^{\mp\mp}$ 
at the LHC and thereby enhancing the possibility of escaping the detection.
 
\end{list}

The investigation of muon $(g-2)$ has given us some useful information about 
the parameters  $y_{\mu\mu}$, $y_{\mu\ell}(\ell \neq \mu)$ and $\mdd$. We are 
 now in a perfect platform to analyse the importance of
these parameters in the context of suitable and relevant CLFV processes, as given
in Table \ref{lfv-limits}. In order to perform this task, we do not
consider the constraint from $(g-2)_\mu$. In this way, we can 
explore the {\it other allowed corner} of the parameter space for $y_{\mu\mu}$ and $y_{\mu\ell}$,
focusing only on the CLFV processes. Subsequently, we will
scrutinize mutual compatibility of the two allowed regions in
$y_{\mu\mu}$ and $y_{\mu\ell}$ parameter space, as obtained from the $(g-2)_\mu$
and CLFV processes. However, to simplify our analysis
we will use one key observation from our discussion of $\Delta a_\mu$,
i.e., in general $y_{\mu\mu}>y_{\mu\ell}$.

The expressions for the branching fractions or the rate of different
CLFV processes are given in Eqs.~(\ref{clfv1}) -- (\ref{mu2e}). From
these formulas it is evident that the allowed region
in $y_{\mu\mu} - y_{\mu\ell}$ parameter space, consistent with
the bounds shown in Table \ref{lfv-limits}, will expand with
larger $\mdd$ values. One can further extract another useful
information from these expressions, i.e., $Br(\tau\to e\gamma)$
and $Br(\tau \to e\mu\mu)$ are the only two CLFV decays without
any $y_{\mu\mu}$ contribution. Both of these processes are
$\propto y^2_{\mu\ell}$ and thus, are in general suppressed compared
to $Br(\tau\to \mu \gamma)$ and $Br(\tau\to 3\mu)$, respectively. 
At the same time, from the view point of present and the expected future
limits (see Table \ref{lfv-limits}), $Br(\tau\to e\gamma)$ $\sim{\cal{O}}$ $(Br(\tau\to \mu\gamma))$
and $Br(\tau\to e\mu\mu)$ $\sim{\cal{O}}$ $(Br(\tau\to 3\mu))$.
Hence, one can safely neglect
the constraints coming from those two processes on the $y_{\mu\mu} - y_{\mu\ell}$
parameter space without any loss of generality. The latter statement
has also been verified numerically.  Thus, we do not consider constraints
from these two channels in our numerical analysis as they will not affect our conclusions
anyway.

\begin{center}
\begin{figure*}
\includegraphics[width=6.55cm]{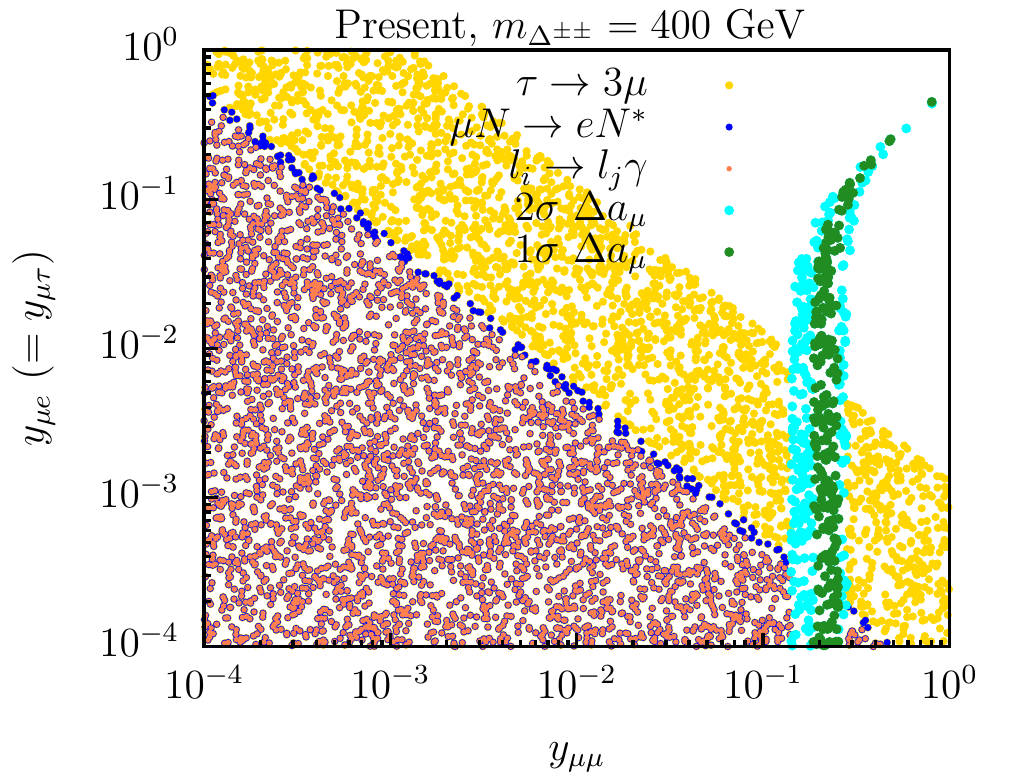}
\includegraphics[width=6.55cm]{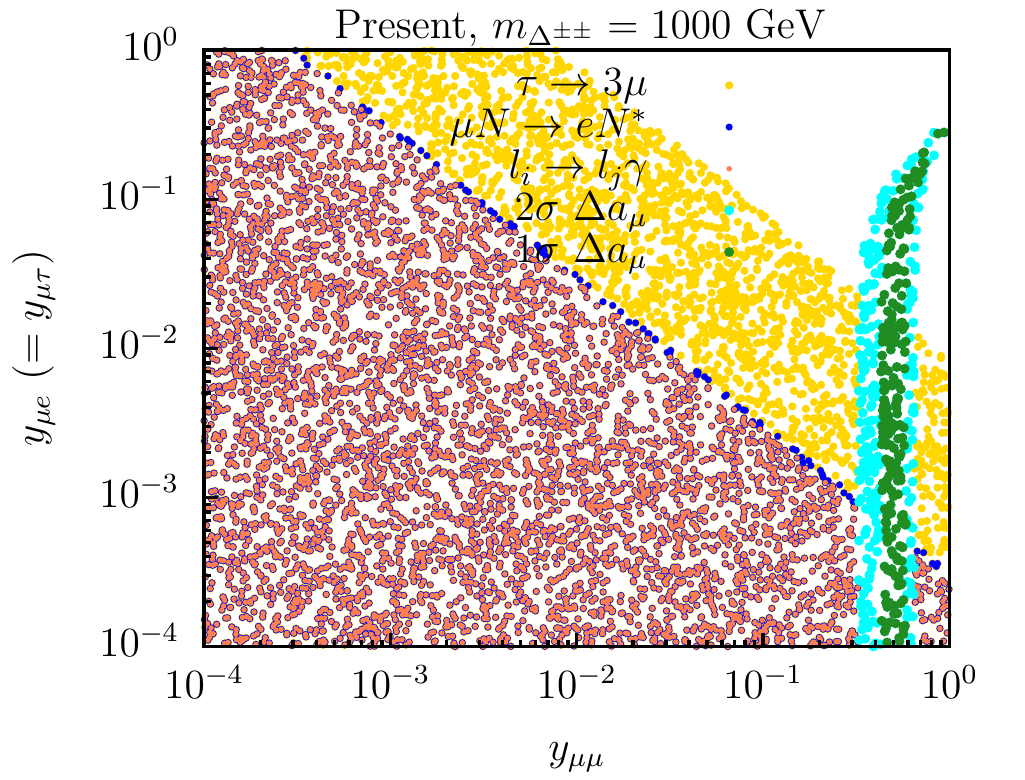}\\
\includegraphics[width=6.55cm]{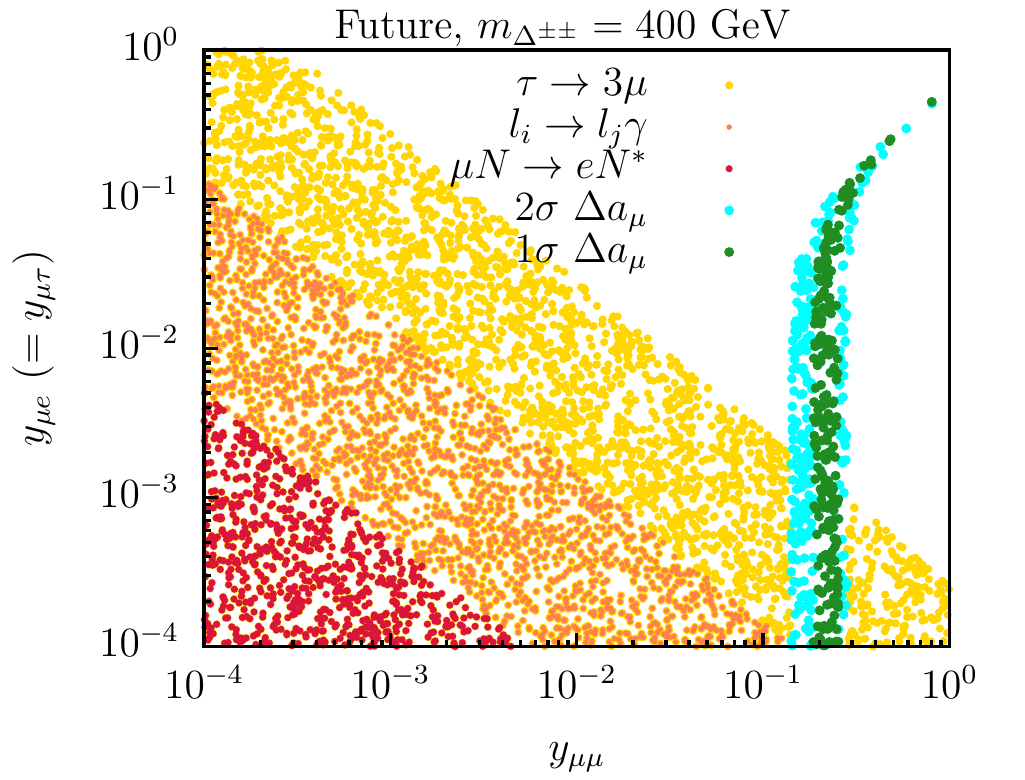}
\includegraphics[width=6.55cm]{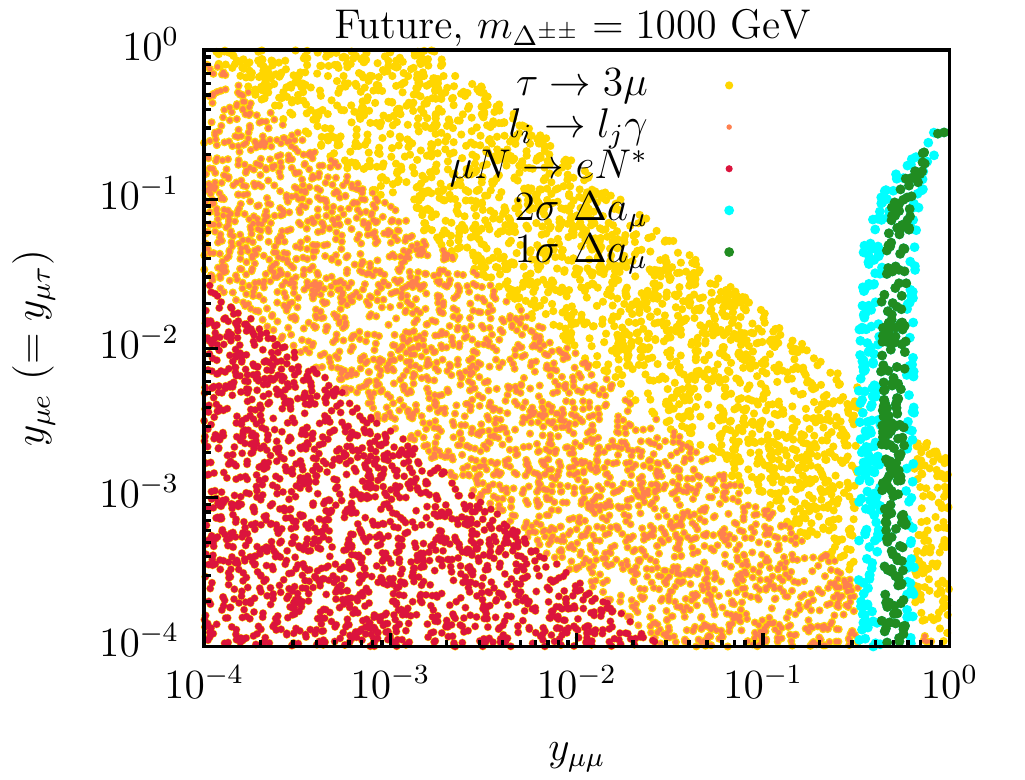}
\caption{\label{fig4n} Plots showing variations of $y_{\mu\mu}$ with
$y_{\mu e}~(\equiv y_{\mu\tau})$ in the context of CLFV processes:
$\mu\to e\gamma$, $\tau \to \mu\gamma$ (collectively phrased as 
$\ell_i\to\ell_j\gamma$), $\tau\to 3\mu$,
$\mu N \to e N^*$ and $\Delta a_\mu$. The plots of
the top row are drawn for $\mdd=400$ GeV (top-left) and $\mdd=1000$ GeV
(top-right) assuming the existing constraints on various CLFV processes (see Table \ref{lfv-limits}).
The sky-blue (deep-green) coloured point represents whether it satisfies the constraint of 
$\Delta a_\mu$ at the $2\sigma~(1\sigma)$ interval.
The golden coloured points (for all the four plots) are those which satisfy the constraint
of only $Br(\tau\to3\mu)$. Deep-blue and dark-red colours (top-row) are used to represent
those points which satisfy the constraint from only $\mu N\to e N^*$ process
and only $\ell_i\to\ell_j \gamma$ process, respectively. The red and orange
coloured points are used to represent the same two quantities in the bottom row plots.
The plots of the bottom row are drawn using the expected
future limits on different allowed CLFV processes.}
\end{figure*}
\end{center}

We plot the allowed region of $y_{\mu\mu} - y_{\mu\ell}$
parameter space in Fig.~\ref{fig4n} using the {\it individual} constraints on 
different CLFV processes as well as on $\Delta a_\mu$, adopting one at a time.
Further, we consider two extreme values of $\mdd$, i.e., $400$ GeV
and $1000$ GeV which cover the entire span. This choice
would help us to understand the relative modification
of the surviving $y_{\mu\mu} - y_{\mu\ell}$
parameter space for a change in $\mdd$ value.
It is clear from all the plots of Fig.~\ref{fig4n} that unlike
$(g-2)_\mu$,   the scales of $y_{\mu\mu}$ and $y_{\mu\ell}$
maintain some kind of reciprocal behaviour. This phenomenon is 
expected since all the formulas of Eqs.~(\ref{clfv1}) -- (\ref{mu2e})
contain the product of Yukawa couplings in the form of  ($y_{\mu\mu} y_{\mu\ell}$). 
The relative arenas
of the allowed $y_{\mu\mu}-y_{\mu\ell}$ regions for 
the different CLFV processes are also well understood. It is apparent
from Table \ref{lfv-limits} that at present the most stringent
limit is coming from $\mu\to e\gamma$, followed
by $\mu N \to e N^*$. On the other hand, the CLFV tau decays have
much larger lower bounds, ${\cal{O}}(10^{-8})$. Hence, as expected,
the allowed $y_{\mu\mu}-y_{\mu\ell}$ parameter space for 
$\mu\to e\gamma$ (thus for $\ell_i\to \ell_j\gamma$) lies in the bottom
(dark-red points in the top-row plots of Fig.~\ref{fig4n}). This region is followed
by the survived $y_{\mu\mu}-y_{\mu\ell}$ parameter space from $\mu N\to e N^*$ process,
since the present limit on $R(\mu N\to e N^*)$ is marginally
larger compared to the present bound on $Br(\mu\to e\gamma)$.
This feature is evident from the narrow visible strip of blue coloured points 
as can be seen in both of the top-row plots of Fig.~\ref{fig4n}. Finally,
rather {\it high} lower limit for $\tau\to 3\mu$ decay leaves a 
large allowed region in the $y_{\mu\mu}-y_{\mu\ell}$ space which
is shown by the golden coloured points. The strip
in the $y_{\mu\mu}-y_{\mu\ell}$ parameter space
which respects the constraint of $\Delta a_\mu$ is very narrow and 
given by the sky-blue (dark-green) coloured points for
the respective $2\sigma~(1\sigma)$ limit [see Eq.~\ref{gm2value}]. 
The presence of $\mdd$ in the denominators [see Eqs.~(\ref{clfv1}) -- (\ref{mu2e})]
suggests an increase of the allowed $y_{\mu\mu}-y_{\mu\ell}$ parameter space
with higher $\mdd$ values. This behaviour is visible
from the two top-row plots of Fig.~\ref{fig4n} where
the surviving $y_{\mu\mu}-y_{\mu\ell}$ region grows larger
for $\mdd=1000$ GeV (top-right plot) compared to
$\mdd=400$ GeV scenario (top-left plot). Independent
study of the allowed CLFV processes and $(g-2)_\mu$
suggests that only a very narrow region of the $y_{\mu\mu}-y_{\mu\ell}$ parameter
space can survive the combined constraints from both.
This region is about $0.15-0.3$ for $y_{\mu\mu}$ while 
$0.0001-0.0004$ for $y_{\mu\ell}$ when $\mdd=400$ GeV (top-left
plot of Fig.~\ref{fig4n}). The span for $y_{\mu\ell}$ increases
slightly, i.e., $0.0001-0.0008$ when one moves to $\mdd=1000$ GeV
(top-right plot of Fig.~\ref{fig4n}). The quantity $y_{\mu\mu}$, 
at the same time, just makes a small shift toward larger values,
i.e., $0.3-0.6$ without expanding the allowed region.

The two plots in the bottom row of Fig.~\ref{fig4n} trail
more or less a similar discussion, especially in 
the context of $\Delta a_\mu$ for which the allowed 
$y_{\mu\mu}-y_{\mu\ell}$ parameter space remains the same. 
This is not true for other processes since these plots
are made using the expected future sensitivities of the
allowed CLFV processes (see Table \ref{lfv-limits}). Now
in the future, the quantity  $R(\mu N\to e N^*)$ is expected
to achieve a lower limit which is about four orders of magnitude smaller
than the current bound. On the contrary, future sensitivities
for $\mu\to e\gamma$ process and CLFV tau decays are only
one order of magnitude smaller than the existing ones.
Thus, in the future the most stringent constraint
on the $y_{\mu\mu}-y_{\mu\ell}$ parameter space
would come from $R(\mu N\to e N^*)$, as shown by the red coloured
points in the two bottom-row plots. The next  most severe
constraint will appear from $\mu\to e\gamma$ (hence for 
$\ell_i\to\ell_j\gamma$) which is represented by orange
coloured points. The golden coloured points represent
the surviving $y_{\mu\mu}-y_{\mu\ell}$ region from the constraint 
of $\tau\to 3\mu$ process. Once again, for each of these concerned processes, 
a larger allowed region in the $y_{\mu\mu}-y_{\mu\ell}$ parameter space appears
as we move from $\mdd=400$ GeV (bottom-left plot) to
$\mdd=1000$ GeV (bottom-right plot). The relative
shrink of the allowed parameter space while using improved 
future bounds, compared to that
with the present constraints, is natural. However,
the important observation from the bottom-row plots  
of Fig.~\ref{fig4n} is the complete disappearance 
of the region of overlap between the surviving $y_{\mu\mu}-y_{\mu\ell}$ parameter
spaces from $R(\mu N\to e N^*)$ and $(g-2)_\mu$. The situation
is practically the same for $Br(\mu\to e \gamma)$ and $(g-2)_\mu$,
although a tiny region of overlap would remain for $\mdd=1000$ GeV (bottom-right plot).
A sizeable region of overlap will still exist between $Br(\tau\to 3\mu)$ and $(g-2)_\mu$ 
processes, however,
smaller compared to the same with present constraints.
So the region of $y_{\mu\mu}-y_{\mu\ell}$ parameter space
that can survive the combined constraints from the possible
CLFV processes and $(g-2)_\mu$ may disappear in the future. This missing area of 
overlap will certainly rule out the possibility of accommodating both the
CLFV processes and $(g-2)_\mu$ in the context of a 
doubly charged scalar in the mass window of $400~{\rm GeV}\lsim \mdd\lsim 1000$ GeV.
Nevertheless, one may observe a region of overlap
like that of the top-row plots with larger values of $\mdd$.
The latter, as already stated, has rather less 
appealing collider phenomenology.

\section{\boldmath $\dc$ at the LHC}\label{collider}

In this final section of our analysis we investigate the collider
phenomenology of a $\dc$ in the light of LHC run-II. 
Our knowledge about the parameters $y_{\mu\mu},\,y_{\mu e}(\equiv y_{\mu\tau})$ and
$\mdd$, as we have acquainted in the last section thus, will appear
resourceful. For the clarity of reading, it is important to 
reemphasise that so far we considered a few low-energy signatures {\it solely} 
from a $\dc$.   Here, we  study the pair-production
of these $\dc$ having hypercharge $Y=1$ at the LHC and so,
for our collider analysis. Thus, the coupling for $\dc\Delta^{\mp\mp}Z_\sigma$ 
vertex (see Sec. \ref{Scenario}) goes\footnote{It is interesting to note that
$\dc\Delta^{\mp\mp}Z_\sigma$ coupling reduces as ones goes from $Y=0$ to $Y=2$. 
For $Y>2$ or for a negative hypercharge, this coupling enhances.}
as $i(g_2 \cos2\theta_W /\cos\theta_W)p^\sigma$. For other choices
of the hypercharge one can simply scale this production
cross-section, $\sigma(p p \xrightarrow{Z/\gamma} \dc\Delta^{\mp\mp})_{Y=1}$ 
as a function of $g_2,\,Y$ and $\sin^2\theta_W$.  Further, we also
assume a negligible/vanishingly small VEV for the {\it possible} neutral scalar component
of this $Y=1$ multiplet and hence, process like $\dc\to W^\pm W^\pm$ becomes
irrelevant. In this scenario, the leading decay modes for a $\dc$ are 
$\ell^\pm_\alpha\ell^\pm_\beta$ which are controlled by $y_{\mu\mu}$
and $y_{\mu e}/y_{\mu\tau}$. It is thus apparent that a set of {\it unconstrained}
$y_{\mu\ell}~(\ell=e,\,\mu,\,\tau)$ couplings will not only produce the same-sign same-flavour
dileptons, e.g., $\dc\to\mu^\pm\mu^\pm$ but will also generate
same-sign different-flavour dileptons, e.g., $\dc\to\mu^\pm e^\pm,\,\mu^\pm\tau^\pm$
with equal branching fractions. The last two decays are example
of lepton flavour violating scalar decays.

At this point, our knowledge of $y_{\mu\mu}$, $y_{\mu e}(\equiv y_{\mu\tau})$
and $\mdd$ from Sec. \ref{result} appears very meaningful to estimate the relative
strengths of different possible $\dc\to \ell^\pm_\alpha\ell^\pm_\beta$ processes.
For our collider analysis, just like our two previous investigations
of a few CLFV processes and $(g-2)_\mu$, we consider
$400~{\rm GeV}\lsim \mdd\lsim 1000$ GeV, following the exclusion 
limit set by the LHC run-I data-set~\cite{ATLAS:2014kca}. At the same
time, from Sec. \ref{result}, we can observe an allowed region in 
the $y_{\mu\mu}-y_{\mu\ell}(\ell\neq \mu)$
parameter space that survives {\it the combined set of present} constraints from
muon $(g-2)$ and a few CLFV processes. In this region of survival, one gets
$y_{\mu\ell}\sim {\mathcal{O}}(10^{-4})$ while $y_{\mu\mu}\sim {\mathcal{O}}(10^{-1})$
(see two top-row plots of Fig.~\ref{fig4n}).
It is hence needless to mention that at the LHC processes
like $\dc\to\mu^\pm e^\pm$ or $\dc\to\mu^\pm\tau^\pm$ will
remain orders of magnitude suppressed compared to
$\dc\to\mu^\pm\mu^\pm$ mode, provided one respects the combined constraints
coming from $(g-2)_\mu$ and a few CLFV processes.
Unfortunately, as discussed in Sec. \ref{result}, such a
conclusion would not hold true in the future when 
the $y_{\mu\mu}-y_{\mu\ell}(\ell\neq \mu)$
parameter space that can survive the combined constraints of
$(g-2)_\mu$ and CLFV processes remains missing. One should note
that such a region in the parameter space can reappear for
large $\mdd$ values, but at the cost of a diminished $\sigma(pp \xrightarrow{Z/\gamma}
\dc\Delta^{\mp\mp})$.
From the aforementioned discussion one can conclude
that with our simplified parameter choice, the region
of parameter space which respects the combined constraints
of muon $(g-2)$ and some CLFV processes will predominantly
yield four-muon ($p p \to\dc\Delta^{\mp\mp}\to 2\mu^\pm2\mu^\mp$) 
final state at the LHC.

For the sake of numerical analyses, the parton level signal events are 
generated using {\tt CalcHEP} \cite{Belyaev:2012qa}. 
These events are then passed through {\tt PYTHIA {v6.4.28}} \cite{Sjostrand:2006za} 
for decay, showering, hadronization, and fragmentation. 
{\tt PYCELL} has been used for the purpose of jet construction.
We have used {\tt CTEQ6L} parton distribution 
function \cite{Pumplin:2002vw} while generating the events. Factorisation and 
renormalisation scales are set at $\sqrt{\hat s}$ (i.e, $\mu_R = \mu_F = \sqrt{\hat s}$), where 
$\sqrt{\hat s}$ is the parton level center-of-mass energy. 
We work in the context of LHC with 13 TeV center-of-mass energy and used the 
following set of {\it basic} selection cuts to identify isolated leptons\footnote{Final 
states with $\tau$-jets (from a hadronically decaying tau) are discarded.}
($l=e,~\mu$) and jets (hadronic) in the final states:

\begin{list}{}{}

\item 
(i)~A final state lepton must have $p_T^{l} > 10$ GeV and $|\eta^{l}| < 2.5$.
\item 
(ii)~A final state jet is selected if $p_T^j > 20$ GeV and $|\eta^j| < 2.5$.
\item 
(iii)~Lepton-lepton separation\footnote{$\Delta R$ is
defined as $\sqrt{(\Delta \Phi)^2+(\Delta \eta)^2}$, where
$\Delta \Phi$ is the difference in involved azimuthal angles while
$\Delta \eta$ is the difference of concerned pseudorapidities, respectively.}, $\Delta R_{l l} > 0.2$. 
\item
(iv)~Lepton-photon separation, $\Delta R_{l\gamma} > 0.2$.
\item 
(v)~Lepton-jet separation, $\Delta R_{l j} > 0.4$.
\item 
(vi)~Hadronic energy deposition, $\sum p^{had}_{T}$ around an isolated lepton must be 
$< 0.2\times p_T^{l}$. 
\item 
(vii)~Final states with four-leptons are selected if 
the leading and subleading leptons have $p_T^{l} > 30$ GeV 
while for the remaining two, $p_T^{l} > 20$ GeV. 
\end{list}

Leading SM background contribution will arise from $t\bar t Z/\gamma^*$ or 
$ZZ/\gamma^*$ events. However, one can use the two following
characteristics to suppress these backgrounds:
(1) Four-lepton final states from $t\bar t Z/\gamma^*$ channels  
always appear with certain amount of missing transverse energy ($\met$).
(2) The set of four-leptons coming from $ZZ$ process contains
pairwise same flavour opposite-sign leptons from a $Z$-decay, and 
 can easily be eliminated using appropriate invariant mass ($m^{inv}$) cuts 
 which is isomorphic to $Z-veto$.
Background events are 
generated using  {\tt MadGraph5@aMCNLO} {v$2.2.3$} \cite{Alwall:2011uj,Alwall:2014hca} and 
subsequently showered with {\tt PYTHIA}. 
In our background simulations, we switched on all the possible processes 
that lead to $p p\to2\mu^\mp2\mu^\pm$ final state with at most two jets
(light or $b$-tagged). At this stage, after a careful scrutiny of the different 
kinematic distributions for both the signal $(S)$ and background $(B)$ events, 
we have introduced the following set of {\it advanced} cuts to guarantee an 
optimized signal to background event ratio:\\
%
{\bf $C1$}: Within the chosen framework a $\dc$ decays only
into $\mu^\pm\ell^\pm~(\ell=e,\,\mu,\,\tau)$ and hence, one would expect no
hadronic jets for the final states. However, hadronic jets may appear while showering 
and we therefore, limit the final state hadronic jet multiplicity 
up to one.\\
{\bf $C2$}: We further impose another criterion on the possible final state 
hadronic-jet, i.e., it must not be a $b$-tagged jet. This choice helps to reduce 
the $t\bar t Z/\gamma^*$ background. \\
{\bf $C3$}: Theoretically, no source of $\met$ 
exists for the predominant decay mode $\dc\to \mu^\pm\mu^\pm$
although nonzero $\met$ can appear from subleading $\dc\to \mu^\pm\tau^\pm$ mode.
The latter, as discussed in the Sec. \ref{result}, remains highly suppressed. 
Hence, we consider an upper limit of 30 GeV on the $\met$.\\
{\bf $C4$}: In our analysis, a pair of same-sign leptons emerges
from a $\dc$ whereas for the backgrounds, a pair of opposite-sign leptons shares the same source.
We therefore, construct $m^{inv}$ for all the possible final state opposite-sign lepton pairs and 
discard all those events with $|m^{inv}_{l^+l^-}-m_Z|\leq 15$ GeV.
Here, $m_Z$ is the mass of $Z$-boson. This cut appears
useful to suppress backgrounds from $Z$-boson decay. 

\begin{table}
\scriptsize
\begin{center}
\begin{tabular}{ c c c c c} 
\hline \hline
\multicolumn{1}{c}{Benchmark} & 
\multicolumn{2}{c}{Parameters}  & 
\multicolumn{1}{c}{Production} &
\multicolumn{1}{c}{Cross-section}
\\
Points & $m_{\Delta^{\pm\pm}}$ (GeV) & $y_{\mu\mu}$ & cross-section & after cuts \\ 
 &  &  & (fb) & (fb) \\ 
\hline  
BP1 & 600 & 0.29  & 1.52 & 0.286 \\
BP2 & 800 & 0.46  & 0.33 & 0.061 \\
BP3 & 1000 & 0.47  & 0.08 & 0.014 \\
\hline
\multicolumn{1}{c}{SM backgrounds} & 
\multicolumn{2}{c}{All inclusive}  & 
\multicolumn{1}{c}{11.56} &
\multicolumn{1}{c}{0.01}
\\
\hline \hline
\end{tabular}
\caption{Signal cross-sections for the three chosen benchmark points before and after 
applying the selection-cuts as described in the text. The final row 
represents the total cross-section from all the possible SM backgrounds 
which contain four-muon final states with a maximum hadronic-jet multiplicity of two.}
\label{signals} 
\end{center}
\end{table}  

In Table~\ref{signals}, we show the signal cross-sections prior and after 
implementing all the basic and advanced cuts. In this context, following the discussion
of last section, we consider a set of three representative benchmark points which simultaneously
satisfies the present set of bounds on CLFV processes and $(g-2)$ of muon. 
The same discussion also predicts $y_{\mu\mu}\gg y_{\mu\ell}(\ell\neq\mu)$ with $y_{\mu\mu}\sim {\cal{O}}(0.1)$
and $y_{\mu\ell}\sim {\cal{O}}(10^{-4})$. Thus, we do not explicitly mention
the corresponding values of $y_{\mu e}(\equiv y_{\mu \tau})$ in Table~\ref{signals}.
In the context of numbers presented in Table~\ref{signals}, it is interesting
to explore the effectiveness of the {\it advanced} cuts, e.g., C3, C4. We have observed
that the advanced selection cut C3 reduces 22\% of the background events
while diminishes 18\% of the signal events (BP1 for example). Subsequent application
of cut C4 kills 4\% of the surviving events for the signal (BP1) whereas
removes 99\% of the surviving background events.

\begin{figure}
\begin{center}
\includegraphics[width=8cm]{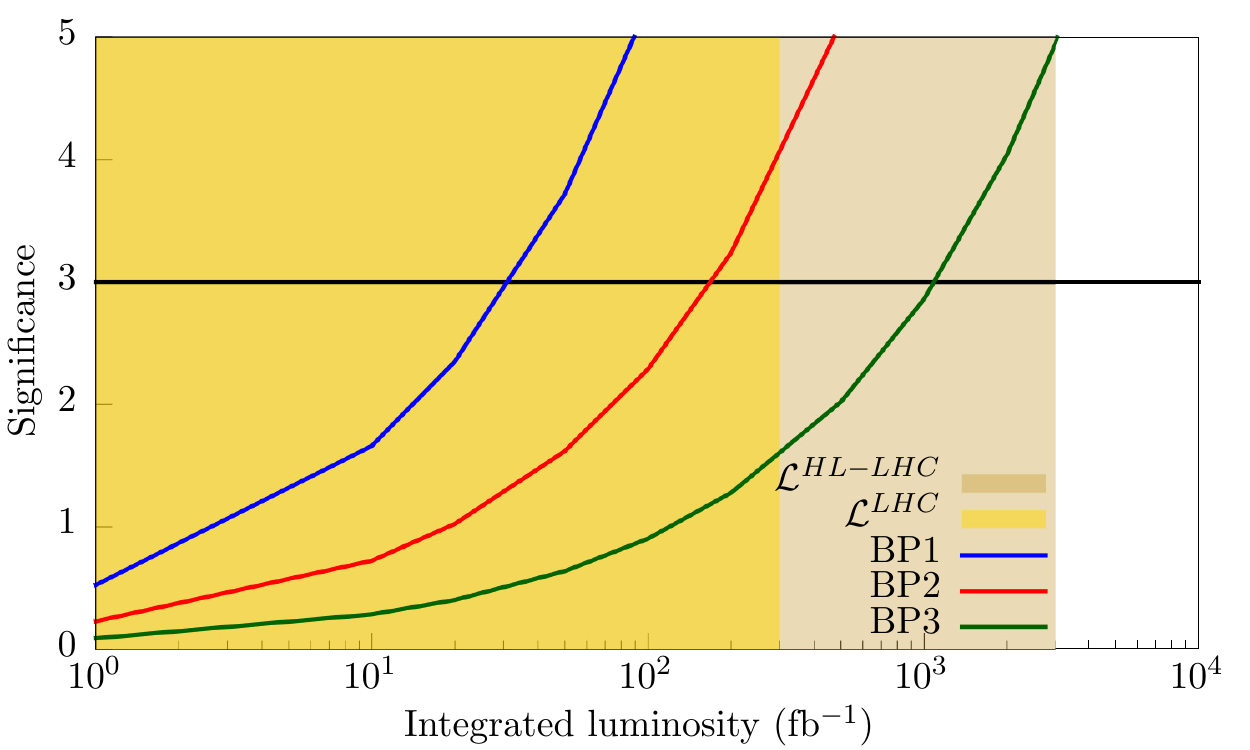}
\caption{Variation of the statistical significance as a function 
of the integrated luminosity $(\mathcal{L})$ for the three different
benchmark points (see Table~\ref{signals}). The black coloured
horizontal line represents a $3\sigma$ statistical significance. 
The golden (dark-golden) coloured band represents the 
luminosity range (with a chosen lower limit
of 1 fb$^{-1}$) for the LHC (proposed high luminosity LHC, HL-LHC).}
\label{fig:sig}
\end{center}
\end{figure}

Finally, in Fig.~\ref{fig:sig} we show the variation of statistical significance\footnote{
Calculated as $S/\sqrt{S+B}$ where $S(B)$ represents the number of 
signal(background) events.} as a function of the integrated luminosity 
$(\mathcal{L})$, for a set of three possible benchmark points (see Table~\ref{signals}).  
The integrated luminosity range (starting from 1 fb$^{-1}$ up to the proposed
maximum) for the LHC and the high-luminosity LHC (HL-LHC) \cite{highlumilhc}) 
are represented with golden and dark-golden colour, respectively. The horizontal black
coloured line represents a $3\sigma$ statistical significance.
The diminishing nature of statistical significance with increasing
$\mdd$ (see benchmark points in Table~\ref{signals})
is a natural consequence of reducing $\sigma (p p \to \dc \Delta^{\mp\mp})$.
One can compensate this reduction with a higher center-of-mass energy
or larger $\mathcal{L}$, as can be seen from Fig.~\ref{fig:sig}. It
is evident from this figure that one would expect strong experimental
evidence (statistical significance $\geq 4\sigma$) of a $\dc$ 
(as sketched within our construction) up to $\mdd\approx 800$ GeV
during the ongoing LHC run-II. The exact time line is however,
$\mdd$ dependent. For example, a discovery (statistical significance $\geq 5\sigma$)
of the studied $\dc$ up to $\mdd\approx 600$ GeV appears
possible with $\mathcal{L}=100$ fb$^{-1}$, which is
well envisaged by 2017 -- 2018. A similar conclusion
for $\mdd=800$ GeV at the $4\sigma$ level however, needs
$\mathcal{L}=300$ fb$^{-1}$ and hence, could appear feasible around 2020.
For a more massive $\dc$ discovery, e.g., $\mdd=1000$ GeV, one would undoubtedly
require a larger $\mathcal{L}$ like $3000$ fb$^{-1}$. Necessity of such a high
$\mathcal{L}$ would leave a massive $\dc$ undetected at the LHC.
The proposed high-luminosity extension of the LHC,
HL-LHC \cite{highlumilhc}, however, will certainly explore this scenario.
We note in passing that a $\dc$ much heavier than $1000$ GeV would remain hidden even 
in such a powerfull machine. The latter, however, will leave its imprints
through a region in the $y_{\mu\mu}-y_{\mu\ell}(\ell\neq\mu)$ parameter
space (see Fig.~\ref{fig4n}) that would simultaneously respect the improved future
bounds on a few CLFV processes and muon $(g-2)$.



\section{Summary and Conclusions}\label{conclusion}

Discovery of a ``Higgs-like'' scalar at the LHC and hitherto 
incomplete knowledge about its origin have revived
the quest for an extended scalar sector beyond the SM.
An interesting possibility is to consider
these extensions through different spin-zero multiplets that
contain various electrically charged (singly, doubly, triply etc.)
and often also neutral fields.
In this paper we have entangled the CLFV and muon (g-2) data to constrain the  relevant parameters associated with  
a doubly charged scalar through a simplified structure and also discuss the possible  collider signatures.
 Further, focusing
on the muon anomalous magnetic moment, we have assumed
only a few nonzero Yukawa couplings, namely $y_{\mu\ell}$
with $\ell=e,\,\mu,\,\tau$, between the doubly
charged scalar and the charged leptons. Furthermore, for simplicity we have chosen
them real as well as $y_{\mu e}=y_{\mu \tau}$ and thus, left with only three relevant free
parameters, namely $y_{\mu\mu},\,y_{\mu e} (\equiv y_{\mu\tau})$ and $\mdd$.

This simplified framework gives two additional contributions to the muon anomalous 
magnetic moment, as shown in Fig.~\ref{fig0}. To start with we have computed
contributions of these two new diagrams in $(g-2)_\mu$ as  
functions of the parameters $y_{\mu\mu},\,y_{\mu e} (\equiv y_{\mu\tau})$ and $\mdd$.
Subsequently, we have scrutinized the impact of individual as well as 
combined contributions from $y_{\mu \mu}$ and $y_{\mu e} (\equiv y_{\mu\tau})$ 
on the $(g-2)_\mu$, for different choices of $\mdd$. We have also
explored various correlations among these three free parameters 
while analysing Figs.~\ref{fig1} -- \ref{fig4}. These correlations
were used to extract the upper bounds on parameters $y_{\mu\mu}$
and $y_{\mu e} (\equiv y_{\mu\tau})$ as $1.2$ and $0.3$, respectively,
keeping in mind their real nature and the span in $\mdd$, i.e., $400~{\rm GeV}$ -- $1000$ GeV.
In addition, these plots also provide the following observations:
(1) Contribution from $y_{\mu e} (\equiv y_{\mu\tau})$ in the evaluation 
of $\Delta a_\mu$ is always negative. 
(2) The size of this contribution is negligible
for $y_{\mu e} (\equiv y_{\mu\tau}) \lsim 0.01$. In
this region, the constraint on $\Delta a_\mu$ gets satisfied
solely from $y_{\mu\mu}$ with $0.1\lsim y_{\mu\mu}\lsim 0.3$. 
(3) In the span of
$0.01\lsim y_{\mu e} (\equiv y_{\mu\tau}) \lsim 0.1$, this contribution
is comparable to the same coming from $y_{\mu\mu}$ 
(i.e., when $0.3\lsim y_{\mu\mu}\lsim 0.6$) and, together
they satisfy the constraint on $\Delta a_\mu$ through a {\it tuned} cancellation.
(4) In the region $0.1\lsim y_{\mu e} (\equiv y_{\mu\tau}) \lsim 0.3$, a
large negative contribution from this parameter appears useful
to compensate the large positive contribution from $y_{\mu\mu}$ with
$y_{\mu\mu}\gsim 0.6$. In this corner of the parameter space, two large but opposite sign
contributions partially cancel each other in a {\it much-tuned way} to satisfy the experimental
bound on $\Delta a_\mu$.

The chosen set of Yukawa couplings also generates new contributions
to a class of CLFV processes, as addressed in 
Sec. \ref{clfv}. We have also investigated these processes
in this paper in the light of parameters $y_{\mu\mu},\,y_{\mu e} (\equiv y_{\mu\tau})$
and $\mdd$, independent of the $(g-2)_\mu$ process. In the context of these analyses
we observed that the allowed $y_{\mu\mu}-y_{\mu\ell}~(\ell\neq \mu)$ parameter
space prefers reciprocal behaviour between the two aforementioned parameters.
This feature is evident from Fig.~\ref{fig4n}. In these same set of plots
we observed a significant enhancement of the surviving $y_{\mu\mu}-y_{\mu\ell}$ parameter
space as one considers larger $\mdd$ values. On the contrary, the allowed
region in the $y_{\mu\mu}-y_{\mu\ell}$ parameter space shrinks 
when one considers more stringent expected future limits on different CLFV processes.
As a final step of our analysis, we have explored the region
of overlap among the different possible $y_{\mu\mu}-y_{\mu\ell}$ planes
that can survive the {\it individual} constraints of various
CLFV processes and $(g-2)_\mu$. Our investigation predicts 
a regime of overlap, i.e., $0.0001\lsim y_{\mu e}(\equiv y_{\mu\tau})\lsim 0.0004$,
$0.1\lsim y_{\mu\mu}\lsim0.3$ for $\mdd=400$ GeV where all the {\it present}
constraints on various CLFV processes and $(g-2)_\mu$ are simultaneously
satisfied. This region, as can be seen from Fig.~\ref{fig4n},
expands slightly for $y_{\mu e}$, i.e.,  $0.0001\lsim y_{\mu e}(\equiv y_{\mu\tau})\lsim 0.0006$
while shifts for $y_{\mu\mu}$, i.e., $0.3\lsim y_{\mu\mu}\lsim0.6$ 
when one considers $\mdd=1000$ GeV.
Expected improvements of the lower bounds for CLFV processes by a few orders of magnitude
in the future, e.g., $R(\mu N\to e N^*)$  would
washout any such common region where constraints
on the CLFV processes and $\Delta a_\mu$ are simultaneously satisfied. Hence, any 
future measurements in this direction will discard the possibility
that {\it only} a doubly charged scalar is instrumental for both the CLFV processes
and the muon anomalous magnetic moment. In other words, 
given that one can achieve the proposed sensitivities for the CLFV
processes in future and observe a region of overlap, the
presence of certain other BSM particles is definitely guaranteed.
One can nevertheless, revive some regime of overlap, even when only a doubly charged scalar 
is present, by considering a much larger $\mdd$ which is experimentally less appealing.

Finally, we used our knowledge of $y_{\mu\mu}$, $y_{\mu e} (\equiv y_{\mu\tau})$
and $\mdd$, that we have gathered while investigating
a few CLFV processes and $\Delta a_\mu$, in the context
of a LHC study for  $pp\to\dc\Delta^{\mp\mp}\to 2\ell^\pm_\alpha 2\ell^\mp_\beta$  processes. 
Our analysis of the Sec. \ref{result} suggests that $y_{\mu\mu}\gg y_{\mu\ell} (\ell\neq \mu)$
when one simultaneously considers the existing set of constraints on the two concerned processes.
Thus, in the context of the chosen simplified
model framework, the decay mode $\dc \to \mu^\pm \mu^\pm$ dominates over
the flavour violating $\dc$ decays. We have addressed
the possibility of detecting our construction at the run-II of LHC
with 13 TeV center-of-mass energy as a function of the integrated luminosity,
for the three different sets of model parameters  (see Fig.~\ref{fig:sig}). 
One can conclude from the same plot that, provided the LHC will attain the proposed
integrated luminosity of 300 fb$^{-1}$, a statistically significant (i.e., $\geq 4\sigma$)
detection of the studied $\dc$ would remain well envisaged
till $\mdd\approx 800$ GeV. Probing higher $\mdd$ values would require
a high-luminosity collider. 
Lastly, we conclude that experimental status of the studied scenario with future generation 
CLFV measurements is rather critical, because:
(1) One observes a region in the $y_{\mu\mu}-y_{\mu\ell}$ parameter space 
which satisfies both the constraints of muon $(g-2)$
and the set of leading CLFV processes for the range of $400~{\rm GeV}\lsim \mdd\lsim 1000$ GeV.
Such an observation would   signify the presence of  some new particles, apart from a $\dc$. 
However, any such additional information will increase the complexity of
the underlying model at the cost of reduced predictability.
(2) A similar region in the $y_{\mu\mu}-y_{\mu\ell}$ parameter
space appears for a higher $\mdd$ value, i.e., $\mdd > 1000$ GeV .
In this case, as can be seen from Fig.~\ref{fig:sig}, the collider
prospects of detecting such a heavy $\mdd$ would appear rather poor, even at
the proposed high luminosity LHC (HL-LHC) with an integrated luminosity of 3000 fb$^{-1}$.


\section*{Appendix}\label{appendA} 

\begin{figure}
\begin{center}
\includegraphics[width=8.0cm,height=1.8cm]{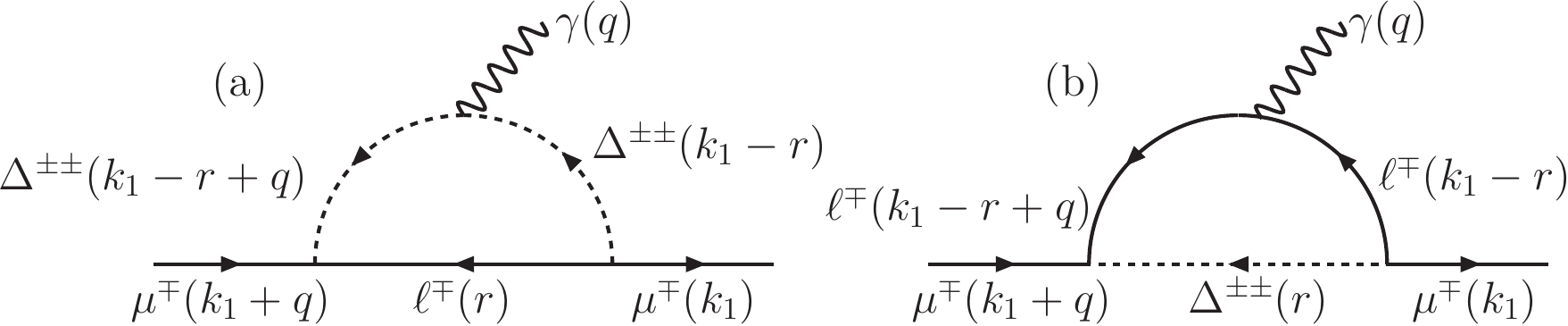}
\caption{\label{fig0n}Relevant details needed to compute Feynman amplitudes 
for the two diagrams shown in Fig.~\ref{fig0}. The direction for all the momentum 
is from left to right
and $k=k_1+q$.}
\end{center}
\end{figure}

In this appendix we present the calculation needed
for the computation of $(g-2)_\mu$ through a $\dc$, as
shown in Fig.~\ref{fig0n}. One may write down these two processes
as $\mu(k_1+q)\to\mu(k_1) + \gamma(q)$, where $k_1,\,q$ represent
 four-momentum of the incoming muon, the outgoing muon, and the outgoing
photon, respectively. 
%

The Feynman amplitude for the process leading to anomalous magnetic moment  of muon can be written as:
\begin{equation}
i \mathcal{M}^\lambda=ie\Big[\bar{u}(k_1 + q)\Big(\gamma^\lambda F_1(q^2) 
+ i \frac{\sigma^{\lambda\nu} q_\nu}{2 m_\mu} F_2(q^2)\Big) u(k_1)\Big],
\end{equation}
where $F_2|_{ q^2=0}$ is  the form factor which needs to be calculated.

The amplitude for the process  shown in Fig.~\ref{fig0n}(a) is written as:
\begin{eqnarray}\label{gm21}
i \mathcal{M}_1^{\dagger} && =  \int \frac{ d^4 r}{(2\pi)^4} \bar{u}(k_1+q) y_{\mu \ell} 
\frac{i(\slashed{r}+m_{\ell})}{r^2-m_{\ell}^2} y_{\mu\ell } \nonumber\\
&& \times \frac{i}{(k_1-r+q)^2-m_{\Delta^{\pm\pm}}^2} \frac{i}{(k_1-r)^2-m_{\Delta^{\pm\pm}}^2}\nonumber\\ 
&& \times \Big[ -iQ_{\Delta^{\pm\pm}}[(k_1-r+q)+k_1-r]_\mu \, \Big]u(k_1),
\end{eqnarray}
where $Q_{\Delta^{\pm\pm}}=2e$ is the  electric charge of the doubly charged scalar.

With the same spirit one can compute the contribution from the second diagram, as shown in 
Fig.~\ref{fig0n}(b), where the amplitude reads as:
\begin{eqnarray}\label{gm22}
i \mathcal{M}_2^{\dagger}&&=\int \frac{ d^4 r}{(2\pi)^4}  \bar{u}(k_1+q) y_{\mu \ell} 
\frac{i(\slashed{k_1}-\slashed{r}+m_{\ell})}{(k_1-r)^2-m_{\ell}^2} (-ie\gamma^\mu) \nonumber \\ 
&&\times \frac{i(\slashed{k_1}-\slashed{r}+\slashed{q}+m_{\ell})}{(k_1-r+q)^2-m_{\ell}^2} 
y_{\mu\ell } \frac{i}{(r^2-m_{\Delta^{\pm\pm}}^2)}  u(k_1).
\end{eqnarray}

After combining these two contributions [Eqs.~(\ref{gm21}), (\ref{gm22})] and extracting the 
coefficient of $\sigma_{\mu \nu}$, after a few intermediate steps, we find the total  
contribution to muon $(g-2)$  as:
\begin{eqnarray}\label{gm2f}
&&\Delta a_\mu =   \frac{f_{m} m_\mu^2 y_{\mu \ell}^2}{8\pi^2} \nonumber\\
&& \times \Bigg[   \int_{0}^{1} 
d\rho\,\,\frac{ 2 (\rho +\frac{m_\ell}{m_\mu})(\rho^2-\rho)}{[m_\mu^2 \rho^2 
+(m_{\Delta^{\pm\pm}}^2-m_\mu^2)\rho+(1-\rho)m_{\ell}^2]}   \nonumber \\
& & - \int_{0}^{1} d\rho\,\,\frac{  (\rho^2-\rho^3 +\frac{m_\ell}{m_\mu} \rho^2)}{[m_\mu^2 \rho^2 
+(m_{\ell}^2-m_\mu^2)\rho+(1-\rho)m_{\Delta^{\pm\pm}}^2]}\Bigg].~
\end{eqnarray}
Here $f_{m}$ is a multiplicative factor which is equal to $1$ for $\ell=e,\,\tau$ 
while equals to $4$ for $\ell=\mu$. The latter appears due to the presence of two identical 
fields in the interaction term.

\vspace{1cm}
\begin{acknowledgments}
The work of J.C. is supported by the Department of Science \& Technology, 
Government of INDIA under the Grant 
Agreement No. IFA12-PH-34 (INSPIRE Faculty Award). 
P.G. acknowledges the support from P2IO Excellence Laboratory (LABEX).
The work of S.M. is partially supported by funding available from 
the Department of Atomic Energy, Government of India, for the Regional 
Centre for Accelerator-based Particle Physics (RECAPP), Harish-Chandra 
Research Institute.
\end{acknowledgments}

\bibliography{Offshell_scalarv7}

\end{document}